\documentclass[%
reprint,
superscriptaddress,
amsmath,amssymb,
aps,
pra,
floatfix
]{revtex4-2}
\usepackage{float}
\usepackage{graphicx}
\usepackage{dcolumn}
\usepackage{bm}
\usepackage{physics}
\usepackage{tabularx}
\usepackage{xcolor} 
\usepackage[caption=false]{subfig} 

\begin{document}

\preprint{APS/123-QED}
\title{Toward Spectral Engineering of Squeezed Light in High-Gain PDC}

\author{Jatin Kumar}
\email{jatin.kumar@uni-jena.de}
\affiliation{Institute of Applied Physics, Abbe Center of Photonics, Friedrich Schiller University Jena, Albert-Einstein-Str. 15, 07745 Jena, Germany}

\author{Aleksa Krsti\'c}
\affiliation{Institute of Applied Physics, Abbe Center of Photonics, Friedrich Schiller University Jena, Albert-Einstein-Str. 15, 07745 Jena, Germany}

\author{Sina Saravi}
\affiliation{Heinz Nixdorf Institute, Paderborn University, Fürstenallee 11, 33102 Paderborn, Germany}
\affiliation{Department of Electrical Engineering and Information Technology, Paderborn University, Warburger Str. 100, 33098 Paderborn, Germany}
\affiliation{Institute for Photonic Quantum Systems (PhoQS), Paderborn University, Warburger Str. 100, 33098 Paderborn, Germany}
\affiliation{Institute of Applied Physics, Abbe Center of Photonics, Friedrich Schiller University Jena, Albert-Einstein-Str. 15, 07745 Jena, Germany}

\author{Frank Setzpfandt} 
\affiliation{Institute of Applied Physics, Abbe Center of Photonics, Friedrich Schiller University Jena, Albert-Einstein-Str. 15, 07745 Jena, Germany}
\affiliation{Fraunhofer Institute for Applied Optics and Precision Engineering IOF, Albert-Einstein-Str. 7, 07745 Jena, Germany}

\date{\today}

\begin{abstract}
{We investigated the spectral properties of squeezed light generated via parametric down-conversion in the high-gain regime, considering both unapodized and apodized dispersion-engineered waveguides. The gain-dependent evolution of these states is examined starting from the low-gain regime, which includes both highly correlated and nearly uncorrelated cases. For the unapodized configuration, we observe a monotonic increase in spectral purity with gain, whereas the apodized configuration exhibits a nonmonotonic dependence, initially decreasing and then recovering at higher gain. By combining Schmidt-mode analysis with a group-velocity-based interpretation, we explain why different dispersion conditions exhibit distinct gain-dependent behavior, specifically that rapid purification occurs when the pump group velocity lies between those of the signal and idler. Our study shows that the evolution of spectral purity is governed primarily by the underlying dispersion of the waveguide. These results demonstrate that dispersion engineering and parametric gain can be jointly exploited to tailor the spectral-mode structure of squeezed-light sources, enabling their optimization for a broad range of quantum applications.}
 
\end{abstract}

\maketitle


\section{Introduction}

Squeezed light has in recent years, become a cornerstone resource in modern quantum technologies, enabling advances in computing, communication, imaging, and metrology \cite{Slussarenko2019,Basset2019,Polino2020, Lawrie2019}. In quantum computing, squeezed states form the basis of continuous-variable (CV) quantum information processing~\cite{Andersen2010,Braunstein2005} and have been used in Gaussian boson sampling to demonstrate quantum computational advantage~\cite{Hamilton2017,Zhong2021,Vernon2019}. In quantum communication, they facilitate secure protocols such as continuous-variable quantum key distribution (CV-QKD) \cite{Madsen2012,Zhang2014,Derkach_2020}. In the domain of quantum imaging, squeezed light improves resolution and sensitivity beyond the classical shot-noise limit, allowing techniques like sub-diffraction imaging and phase-contrast microscopy~\cite{Kolobov1993,Brida2010,Taylor2014}. Finally, in quantum metrology, squeezed states enhance precision by reducing quantum noise in interferometric measurements \cite{Caves1981,Giovannetti2011}, most notably improving the sensitivity of gravitational wave detectors such as LIGO via squeezed vacuum injection  \cite{Aasi2013}.


Squeezed light is predominantly generated via nonlinear optical processes \cite{Slusher,Wu,Shelby}, with parametric down-conversion (PDC) being a widely employed technique \cite{Loudon}.  In PDC, pump photons propagating through a $\chi^{(2)}$ nonlinear medium are converted into pairs of signal and idler photons. This process obeys the laws of energy and momentum conservation, leading to quantum correlations between the resulting signal and idler photons \cite{Bennink2006}. At sufficiently low pump intensities, mainly signal and idler photon pairs are generated, and the PDC source is said to operate in the low-gain regime. In contrast, when a high-intensity pump is used, the generated photon pairs can further stimulate down-conversion events, leading to an exponential increase in photon production; this regime is known as the high-gain regime \cite{Brambilla}. Under these conditions, the quantum correlations of individual photon pairs collectively give rise to strong squeezing~\cite{Sharapova2020}.

For practical applications, precise control over the spectral, temporal, and statistical properties of squeezed light is crucial \cite{Andersen2016,Lvovsky2009,TAYLOR20161}.
The advent of integrated photonics platforms has opened new possibilities to control these properties to a very high degree \cite{Lenzini2018}.
In particular, waveguides can be engineered to precisely control effective indices, group velocities, and phase-matching conditions \cite{Kogelnik,Silverstone,Tanzilli}, thereby tailoring the spectral and temporal properties of the generated quantum fields. As a result, integrated waveguide platforms have emerged as ideal sources for the generation of quantum light, offering enhanced scalability, stability, and control \cite{Wang2018}. Lithium niobate on insulator (LNOI) is an exceptionally suitable platform for integrated quantum photonics due to its strong $\chi^{(2)}$ nonlinearity, tight modal confinement, low propagation loss, and flexible dispersion engineering \cite{Wang2018,Krasnokutska2018,Zhu2021}. In addition, the efficiency of nonlinear interactions can be further enhanced via poling \cite{Chen2024}. Several experimental demonstrations on LNOI-based platforms have successfully generated squeezed light, including broadband squeezed vacuum generation \cite{Chen2022}, squeezing at telecom wavelengths \cite{Peace2022}, and quadrature squeezed light in integrated nanophotonic waveguides \cite{Park2024}.


When designing a high-gain source, in addition to the aforementioned platform-specific techniques, special care must be taken to account for the effects that high-gain operation has on the spectral, temporal, and statistical properties of the output light. These effects include spectral broadening, changes in spectral purity, temporal reshaping, and photon bunching, each of which can significantly impact the performance of quantum applications \cite{Spasibko2012, Christ2011, Sharapova2018, Iskhakov2012}. Understanding these high-gain effects is therefore crucial for optimizing source performance in practical implementations. 

In this work, our goal is to understand how the spectral correlations in the output state of a waveguide-based PDC source are affected by increasing gain, and whether certain types of low-gain correlations lead to more favorable spectral properties in the high-gain regime. More broadly, we aim to clarify, in principle, how different low gain spectral configurations behave under high-gain conditions and what implications they carry for the engineering of quantum light sources tailored to specific application demands.

In this paper, we employ a theoretical formalism for high-gain parametric down-conversion (PDC) in waveguides that accommodates arbitrary dispersion relations. This framework is particularly well-suited for modeling modern integrated platforms, such as lithium niobate on insulator (LNOI), where complex dispersion can be harnessed to engineer quantum state generation \cite{Wang2017Metasurface, Fang2020Second}. We investigate the key waveguide parameters that control the spectral properties of the generated squeezed light. Specifically, we focus on three sets of dispersion conditions to study the behaviour of gain and spectral purity for apodized and unapodized phasematching scenarios. In Section II, the theoretical formalism is discussed. In Section III, we investigate high-gain PDC through numerical simulations and analyze its gain-dependent spectral behavior. Finally, conclusions are discussed at the end.

\section{Theory}\label{sec:theory}

The theoretical description of nonlinear light–matter interactions depends strongly on how dispersion is treated. In many works, a weak-dispersion approximation is used, where the temporal degree of freedom is replaced by a spatial one \cite{Quesada, Houde2023}. In this work we adopt a time-domain approach that accurately accounts for arbitrary dispersion. This more general formalism enables us to capture complex dispersion effects that can arise in waveguides engineered with nanostructuring or other intricate geometrical features \cite{Wang2023,Cai2018}.

\begin{figure}[!htpb]
\includegraphics[width=0.45\textwidth, keepaspectratio]{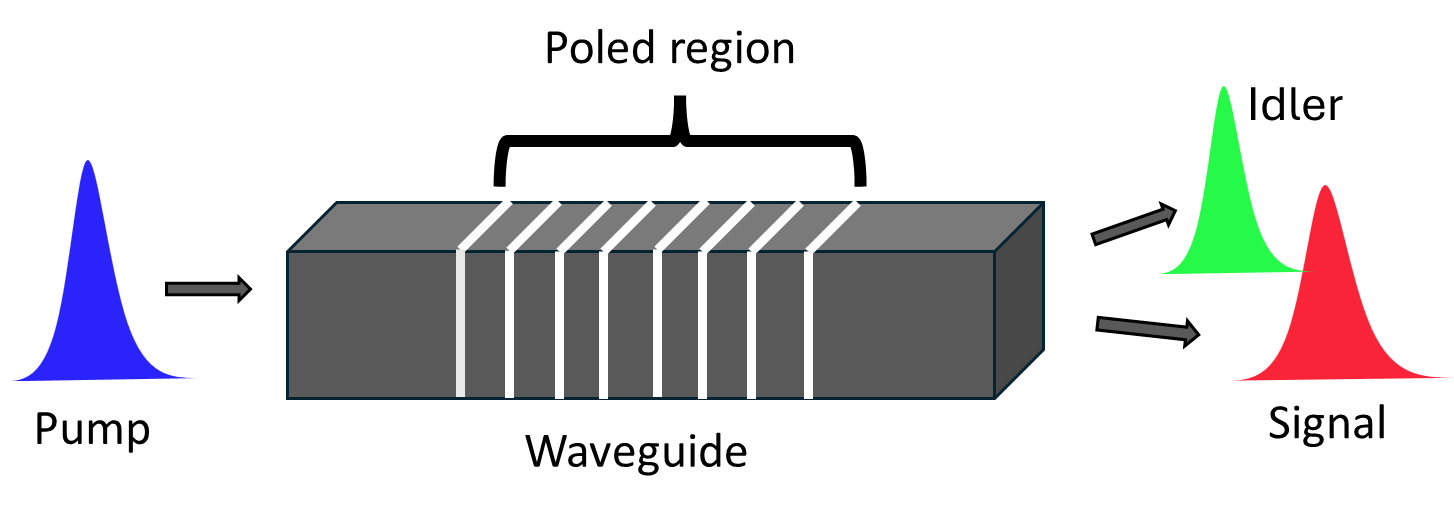}
\caption{Schematic of PDC process in a waveguide where pump interacts with the nonlinear medium of the waveguide to generate signal and idler fields.}
\label{fig:wg}
\end{figure}

Here, we consider a nonlinear waveguide with pulsed pump to generate two-mode squeezed light, where the signal and idler modes are distinct guided modes and all modes propagate in the same direction. The waveguide is assumed to be infinitely extended along the $z$-axis, with a finite poled region of length $L$, as illustrated in Fig.~\ref{fig:wg}. The transverse geometry, together with the dielectric permittivity distribution, determines the linear characteristics of the guided modes, specifically their effective indices, group velocities, and transverse mode profiles.

For the aforementioned waveguide, assuming that material dispersion is negligible, the quantized electric field for the signal and idler modes (positive frequency part) is defined as \cite{Yang2008, Andreas}:
\begin{equation}
	\begin{aligned}
		\hat{\mathbf{E}}^{j(+)}(\mathbf{r},t) &= i\gamma\int\dd{\omega} \sqrt{\frac{\omega n^{j}_g(\omega)}{ N_{j}(\omega)} }\,
		\textbf{E}^{j}(x,y,\omega) \\
		&\quad\times \exp{i(k_{j}(\omega)z)} \hat{c}^{j}_{\omega}(t), \quad j \in \{s,i\},
	\end{aligned}
\end{equation}
where \(j= s\) corresponds to the signal mode and \(j=i\) corresponds to the idler mode, \(\gamma = \sqrt{\frac{\hbar }{4\pi \varepsilon_0 c}}\), \(\textbf{E}^{j}(x,y,\omega)\) is the transverse mode profile, \(n^{j}_g(\omega)\) is the modal group index, and $N_{j}(\omega) = \iint \mathrm{d}x\,\mathrm{d}y\, \varepsilon_r(x,y,\omega) |\mathbf{E}^{j}(x,y,\omega)|^2
$ is a mode normalization constant. Here, \(\hat{c}^{s}_{\omega}(t) \equiv \hat{a}_\omega(t)\) and \(\hat{c}^{i}_{\omega}(t) \equiv \hat{b}_\omega(t)\) are the annihilation operators for the signal and idler modes, respectively.
The annihilation operators $\hat{c}^{j}_{\omega}(t)$ and creation operators $\hat{c}^{j'\dagger}_{\omega}(t)$ obey the canonical commutation relation
$[\hat{c}^{j}_{\omega}(t),\hat{c}^{j'\dagger}_{\omega'}(t)] = \delta_{jj'} \delta(\omega-\omega')$.

In this work, we employ the Heisenberg picture. In addition, throughout the paper, spatial integrals are taken over the transverse cross-sectional area $A$ and along the longitudinal axis of the waveguide over the domain $z \in [0, L]$; frequency integrals are taken over the interval $[0, \infty)$, and time-domain integrals extend over $(-\infty, \infty)$.


To describe the parametric down-conversion (PDC) process, we consider the following form of the nonlinear interaction Hamiltonian \cite{Andreas},
\begin{equation}
	\begin{aligned}
		\hat{H}_{\mathrm{PDC}}(t) 
		&= -2\varepsilon_0 \int \dd^{3}\mathbf{r}\,
		\chi_{ljm}^{(2)}(\mathbf{r})\,
		\mathrm{E}^{p(+)}_{l}(\mathbf{r},t)\,
		\hat{\mathrm{E}}^{s(-)}_{j}(\mathbf{r},t)\,\\ &\quad\times
		\hat{\mathrm{E}}^{i(-)}_{m}(\mathbf{r},t)
		+ \text{H.c.}.
	\end{aligned}
\end{equation}

Here, $\varepsilon_0$ is the vacuum permittivity, $\chi^{(2)}_{ljm}$ is the second-order nonlinear susceptibility tensor, $\mathrm{E}^{p (+)}_{l}(\mathbf{r},t)$ is the positive-frequency component of the pump field, $\hat{\mathrm{E}}^{s (-)}_{j}(\mathbf{r},t)$ is the negative-frequency component of the signal field, $\hat{\mathrm{E}}^{i (-)}_{m}(\mathbf{r},t)$ is the negative-frequency component of the idler field, and $\text{H.c.}$ denotes the Hermitian conjugate.
To drive the nonlinear interaction in the waveguide, we assume a classical, undepleted pump pulse with a Gaussian temporal envelope.

The Gaussian pump pulse is defined as:
\begin{equation}
\begin{aligned}
    \mathbf{E}^{p(+)}(\mathbf{r},t) &= 
    \gamma_{p} \int \dd{\omega_{p}} \, \sqrt{\frac{n^{p}_{g}(\omega_{p_0})}{N_{p}(\omega_{p})}}\alpha(\omega_{p}) \mathbf{E}^{p}(x,y,\omega_{p}) \\
    & \quad \times \exp{i(k_{p}(\omega_{p}) z - \omega_{p} t)},
    \label{eq:pump}
\end{aligned}
\end{equation}
where $\gamma_p = \sqrt{\frac{\mathrm{P}_p }{4\pi\varepsilon_0 c\sigma^{2}_{p} }}$ incorporates the pump peak power $\mathrm{P}_p$, speed of light $c$, and vacuum permittivity $\varepsilon_0$. The group index is denoted by $n^{p}_{g}$. The function $\alpha(\omega_p) = \exp\left(-\frac{(\omega_p-\omega_{p_0})^2}{2\sigma_p^2}\right)$ represents the Gaussian spectral envelope centered at \(\omega_{p_0}\) with bandwidth \(\sigma_p\). \(\mathbf{E}^{p}(x,y,\omega_p)\) is the transverse field profile of the pump; and \(N_{p}(\omega_p) = \iint \dd{x}\,\dd{y}\, \varepsilon_r(x,y,\omega_p) |\mathbf{E}^{p}(x,y,\omega_p)|^2\) is the normalization constant for the transverse pump mode profile with relative permittivity \(\varepsilon_r\).

Since we are working in the Heisenberg picture, the total Hamiltonian of the system includes the free (time-independent) part \(H_0\) and the nonlinear interaction part \(H_{\mathrm{PDC}}(t)\). The free Hamiltonian is
\begin{equation}
H_0 = \sum_{j=s,i}\int \mathrm{d}\omega_j \,\hbar\omega_j\,\hat{c}^{j\dagger}_{\omega}(t)\hat{c}^{j}_{\omega}(t),    
\end{equation}
which governs the free evolution of each mode. To remove this trivial evolution we transform to a rotating frame by
\[
\hat{c}^{j}_{\omega}(t)=\tilde{c}^{j}_{\omega}(t)\,e^{-i\omega t},\qquad j\in\{s,i\},
\]
where \(\tilde{c}^{j}_{\omega}(t)\) are slowly varying operators that evolve only due to the nonlinear interaction \cite{gerry2005introductory,Alexa}.  With this transformation, the Heisenberg equation of motion for the slowly varying operators becomes
\begin{equation}
 i\hbar\partial_t\tilde{c}^{j}_{\omega}(t) = \big[\tilde{c}^{j}_{\omega}(t),\,H_{\mathrm{PDC}}(t)\big].  
\end{equation}
Applying this to the slowly varying signal and idler operators, $\Tilde{a}_{\omega^{'}_s}(t)$ and $\Tilde{b}_{\omega^{'}_i}(t)$, we obtain coupled integro-differential equations among the creation and annihilation operators
\begin{equation}
\begin{aligned}
\partial_{t}\Tilde{a}_{\omega^{'}_s}(t) &= -\frac{i}{\hbar}\int\dd{\omega_{i}}G_{1}(\omega_i,\omega^{'}_s,t)\Tilde{b}^{\dagger}_{\omega_{i}},\\
\partial_{t}\Tilde{b}_{\omega^{'}_i}(t) &= -\frac{i}{\hbar}\int\dd{\omega_{s}}G_{2}(\omega_s,\omega^{'}_i,t)\Tilde{a}^{\dagger}_{\omega_{s}}.
\label{eq:heisenberg_ab}
\end{aligned}
\end{equation}

These coupled integro-differential equations involve the kernels $G_{1}(\omega_i,\omega^{'}_s,t)$ and $G_{2}(\omega_s,\omega^{'}_i,t)$, which represent the nonlinear coupling between the modes. The kernels $G_{1}(\omega_i,\omega^{'}_s,t)$ and $G_{2}(\omega_s,\omega^{'}_i,t)$ are defined as:
\begin{widetext}
\begin{equation}
    G_1(\omega_i,\omega^{'}_s,t) = M \frac{\hbar \gamma_p}{2\pi c}
    \sqrt{\omega^{'}_s \omega_{i} n^{p}_g(\omega_{p_0})n^{s}_g(\omega^{'}_s) n^{i}_g(\omega_{i})}
    \int\dd{\omega_{p}}\alpha(\omega_{p}) \phi(\omega_i,\omega^{'}_s,\omega_p) 
\exp\left[-i(\omega_{p}-\omega_{i}-\omega^{'}_s)t\right],
\label{eq:g1}
\end{equation}
\begin{equation}
    G_2(\omega_s,\omega^{'}_i,t) = M\frac{\hbar \gamma_p}{2\pi c}\sqrt{\omega^{'}_i \omega_{s} n^{p}_g(\omega_{p_0})n^{i}_g(\omega^{'}_i) n^{s}_g(\omega_{s})}
\int\dd{\omega_{p}}\alpha(\omega_{p})\phi(\omega_s,\omega^{'}_i,\omega_p)
    \exp\left[-i(\omega_{p}-\omega_{s}-\omega^{'}_i)t\right].
\label{eq:g2}
\end{equation}
\end{widetext}
The key factors in the coupling kernels are the phasematching function $\phi(\omega_i,\omega_s,\omega_p)$ and the nonlinear mode-overlap factor $M$. The phasematching function is defined as
\begin{equation}
    \phi(\omega_i,\omega_s,\omega_p) 
    = \int \dd{z}\, f(z)\, 
    \exp\!\big[i\,\Delta k(\omega_i,\omega_s,\omega_p)\,z\big],
    \label{eq:phasematching_function}
\end{equation}
where $f(z)$ represents the spatial modulation of the nonlinearity along the propagation direction, determined by the poling profile of the waveguide (see Appendix~\ref{app:phasematching} for details). The phase mismatch between the interacting modes is defined as:
\begin{equation}
    \Delta k(\omega_i,\omega_s,\omega_p) =
    k_{p}(\omega_{p}) - k_{s}(\omega_s) - k_{i}(\omega_{i}).
    \label{eq:delK}
\end{equation}
Here $k_p(\omega_p)$, $k_s(\omega_s)$, and $k_i(\omega_i)$ are the propagation constants of the pump, signal, and idler modes, respectively. The nonlinear mode-overlap factor $M$ is given by:
\begin{equation}
\begin{aligned}
 M &= \int \dd{x}\,\dd{y}\,
 \frac{\chi^{(2)}_{ljm}}{\sqrt{N_{p} N_s N_i}}\,
 \mathrm{E}^{p}_{l}(x,y,\omega_p)\,
 \mathrm{E}^{s*}_{j}(x,y,\omega_s)\,\\
 &\quad\times\mathrm{E}^{i*}_{m}(x,y,\omega_i),
\label{eq:M}    
\end{aligned}
\end{equation}
and is evaluated under the assumption that the transverse mode profiles remain effectively constant over the frequency range of interest.

Since the PDC Hamiltonian is quadratic in the creation and annihilation operators, we can use a Bogoliubov transformation to describe the evolution of the operators \cite{NAM20164340}. This transformation, also known as the input-output transformation, is a standard theoretical tool for describing high-gain parametric down-conversion (PDC) \cite{Helt, Andreas}. Here, the input operators correspond to the initial time and the output operators correspond to the final time. The input-output relations for $\Tilde{a}_{\omega_s}(t)$ and $\Tilde{b}_{\omega_i}(t)$ are expressed as
\begin{equation}
\begin{split}
    \Tilde{a}_{\omega_s}(t) &= \int \dd{\omega^{''}_s} A(\omega_s,\omega^{''}_s,t)\Tilde{a}_{\omega^{''}_s}(t_0) \\
    &\quad+ \int \dd{\omega^{''}_i} B(\omega_s,\omega^{''}_i,t)\Tilde{b}^{\dagger}_{\omega^{''}_i}(t_0),
\label{eq:inouta}
\end{split}
\end{equation}

\begin{equation}
\begin{split}
    \Tilde{b}_{\omega_i}(t) &= \int \dd{\omega^{''}_i}C(\omega_i,\omega^{''}_i,t)\Tilde{b}_{\omega^{''}_i}(t_0)\\&\quad+
\int \dd{\omega^{''}_s}D(\omega_i,\omega^{''}_s,t)\Tilde{a}^{\dagger}_{\omega^{''}_s}(t_0),
\label{eq:inoutb}
\end{split}
\end{equation}
where $A(\omega_s,\omega''_s,t)$, $B(\omega_s,\omega''_i,t)$, $C(\omega_i,\omega''_i,t)$, and $D(\omega_i,\omega''_s,t)$ are complex-valued scalar functions that relate the input and output operators. To ensure that the Bogoliubov transformation preserves the canonical commutation relations, the complex-valued scalar functions must satisfy specific constraints. These constraints provide a necessary consistency check of the formalism and are 
\begin{equation}
\begin{aligned}
[\tilde{a}_{\omega_s}(t), \tilde{a}^\dagger_{\omega_s'}(t)] &= 
\int \mathrm{d}\omega_s''\, A(\omega_s, \omega_s'', t) A^*(\omega_s', \omega_s'', t) \\
&\quad - \int \mathrm{d}\omega_i''\, B(\omega_s, \omega_i'', t) B^*(\omega_s', \omega_i'', t) \\
&= \delta(\omega_s - \omega_s'),
\end{aligned}
\end{equation}
\begin{equation}
\begin{aligned}
[\tilde{b}_{\omega_i}(t), \tilde{b}^\dagger_{\omega_i'}(t)] &= 
\int \mathrm{d}\omega_i''\, C(\omega_i, \omega_i'', t) C^*(\omega_i', \omega_i'', t) \\
&\quad - \int \mathrm{d}\omega_s''\, D(\omega_i, \omega_s'', t) D^*(\omega_i', \omega_s'', t) \\
&= \delta(\omega_i - \omega_i'),
\end{aligned}
\end{equation}
\begin{equation}
\begin{aligned}
[\tilde{a}_{\omega_s}(t), \tilde{b}_{\omega_i}(t)] &= 
\int \mathrm{d}\omega_s''\, A(\omega_s, \omega_s'', t) D(\omega_i, \omega_s'', t) \\
&\quad - \int \mathrm{d}\omega_i''\, B(\omega_s, \omega_i'', t) C(\omega_i, \omega_i'', t) \\
&= 0.
\end{aligned}
\end{equation}





Substituting equations (\ref{eq:inouta}) and (\ref{eq:inoutb}) into equation (\ref{eq:heisenberg_ab}), we obtain a set of coupled integro-differential equations expressed in terms of complex scalar functions. With the infusion of the Bogoliubov transformation in the operator formalism, the description transitions to a scalar formalism, which is easier to handle and allows us to work directly with these complex valued functions A, B, C, and D. The resulting coupled integro-differential equations are given by

\begin{equation} 
\begin{split}
 \partial_{t}A(\omega^{'}_s,\omega^{''}_s,t) &= -\frac{i}{\hbar}\int\dd{\omega_{i}}G_{1}(\omega_i,\omega^{'}_s,t)D^{*}(\omega_i,\omega^{''}_s,t),\\
\partial_{t}B(\omega^{'}_s,\omega^{''}_i,t)&=  -\frac{i}{\hbar}\int\dd{\omega_{i}}G_{1}(\omega_i,\omega^{'}_s,t)C^{*}(\omega_i,\omega^{''}_i,t),\\
\partial_{t}C(\omega^{'}_i,\omega^{''}_i,t) &= -\frac{i}{\hbar}\int\dd{\omega_{s}}G_{2}(\omega_s,\omega^{'}_i,t)B^{*}(\omega_s,\omega^{''}_i,t),\\
\partial_{t}D(\omega^{'}_i,\omega^{''}_s,t) &= -\frac{i}{\hbar}\int\dd{\omega_{s}}G_{2}(\omega_s,\omega^{'}_i,t)A^{*}(\omega_s,\omega^{''}_s,t).
\label{eq:coupledeq}
\end{split}
\end{equation}

Since all the field properties are embedded in the complex functions, we study the properties of the squeezed state by solving the resulting coupled integro-differential equations. We solve these equations numerically by incorporating the initial conditions

\begin{equation} 
\begin{split}
A(\omega^{'}_s,\omega^{''}_s,t_0) & = \delta(\omega^{'}_s-\omega^{''}_s),\\
C(\omega^{'}_i,\omega^{''}_i,t_0) & =\delta(\omega^{'}_i-\omega^{''}_i),\\
D(\omega^{'}_i,\omega^{''}_s,t_0)&= 0,\\
B(\omega^{'}_s,\omega^{''}_i,t_0)&=0.
\end{split}
\end{equation}

These initial conditions are determined from Eqs. (\ref{eq:inouta}, \ref{eq:inoutb}) evaluated at time $t_0$. 

Since squeezed states belong to the class of Gaussian states, which are fully characterized by their first- and second-order moments \cite{Weedbrook}, the output quantum state produced by the nonlinear waveguide can be completely described using these moments. In our case, the first-order moments of the form $\expval{\Tilde{a}^{(\dagger)}_{\omega}(t)}$ and $\expval{\Tilde{b}^{(\dagger)}_{\omega}(t)}$ vanish, so only the second-order moments need to be calculated, which are given by:

\begin{equation}
\begin{split}
\bra{0}\Tilde{a}_{\omega_s}(t)\Tilde{b}_{\omega_i}(t)\ket{0}&= \int \dd{\omega_{s}^{''}}A(\omega_s,\omega^{''}_s,t)D(\omega_i,\omega^{''}_s,t),
\label{eq:spectral_corr}
\end{split}
\end{equation}

To investigate the spectral properties of the output quantum state, we further decompose the second-order moments defined in Eq.~(\ref{eq:spectral_corr}) in terms of the Schmidt modes. These modes are obtained via the singular value decomposition (SVD) of the scalar functions from Eq.~(\ref{eq:inouta}) and Eq.~(\ref{eq:inoutb}) \cite{Christ2011}.
The Schmidt mode decomposition provides a clear interpretation of the multimode structure of squeezed light in the spectral domain. It offers an intuitive physical picture where each mode corresponds to a distinct wavepacket \cite{Law2000}, allowing the signal and idler beams to be described as superpositions of these modes.
In terms of the Schmidt modes, the second-order moments are expressed as \cite{Houde2023}

\begin{equation}
\begin{split}
\bra{0}\Tilde{a}_{\omega_s}(t)\Tilde{b}_{\omega_i}(t)\ket{0} &=\sum_{\ell}\frac{\sinh(2r^{(\ell)})}{2}\psi^{(\ell)}_s(\omega_s)\psi^{(\ell)}_i(\omega_i),
\label{eq:spectral_corr_schmidt}
\end{split}
\end{equation}

where the Schmidt modes, denoted as $\psi^{(\ell)}_{s/i}$ for the signal and idler respectively, form an orthonormal basis. Each pair of modes $\ell$ is associated with a squeezing parameter $r^{(\ell)}$, which quantifies the squeezing strength in that mode. 
From the second-order moment in Eq.~(\ref{eq:spectral_corr_schmidt}), expressed in terms of Schmidt modes, we can reconstruct the joint spectral amplitude (JSA). The JSA is given by
\begin{equation}
    \mathrm{JSA}(\omega_s, \omega_i) = \sum_{\ell}r^{(\ell)}\psi^{(\ell)}_s(\omega_s)\psi^{(\ell)}_i(\omega_i).
\end{equation}
In the low-gain regime, where $r^{(\ell)}<<1$, the second-order moment in Eq.~(\ref{eq:spectral_corr_schmidt}) converges to the JSA, so no reconstruction is needed. However, at high gain (large $r^{(\ell)}$) the second-order moment no longer directly represents the JSA, and explicit reconstruction is required. 

Since our goal is to characterize the spectral properties of squeezed states relevant for practical applications, the spectral purity provides a physically meaningful figure of merit. To evaluate it, we extract the squeezing parameters $r^{(\ell)}$ from Eq.~(\ref{eq:spectral_corr_schmidt}), which encodes the Schmidt-mode structure of the state. The spectral purity $\mathcal{P}$ is defined as \cite{Houde2023}
\begin{equation}
	\mathcal{P} = \frac{\sum_{\ell}\sinh^{4}\!\big(r^{(\ell)}\big)}
	{ \left(\sum_{\ell}\sinh^{2}\!\big(r^{(\ell)}\big)\right)^2 } .
	\label{eq:purity}
\end{equation}
It provides a quantitative measure of the multimode character of the squeezed state, with $\mathcal{P}=1$ corresponding to an ideal single-mode squeezed state.

In addition, the extracted squeezing parameters \( r^{(\ell)} \) quantify the overall strength of the parametric process. Following Ref.~\cite{Christ2011}, we define an overall
parametric gain \( G \) as
\begin{equation}
	G = \sqrt{\sum_{\ell} \big(r^{(\ell)}\big)^2 } .
	\label{eq:G_def}
\end{equation}
The corresponding gain expressed in decibels is given by
\begin{equation}
	G_{\mathrm{dB}} = -10\log_{10}\!\big(e^{-2G}\big).
	\label{eq:GdB_def}
\end{equation}
Finally, to study the gain-dependent behavior of the Schmidt modes, we perform a mode-contribution analysis. For this purpose, we introduce the normalized
modal weights \( p_\ell \), defined as
\begin{equation}
	p_{\ell}
	= \frac{\sinh^{2}\!\big(r^{(\ell)}\big)}
	{\sum_{\ell}\sinh^{2}\!\big(r^{(\ell)}\big)} ,
	\qquad
	\sum_{\ell} p_{\ell} = 1 .
	\label{eq:p_def}
\end{equation}
These weights quantify the fractional photon-pair population associated with the
\(\ell\)-th Schmidt mode pair~\cite{Houde2023}.

\section{Simulation}
\label{sec:simulation}
In this section, we investigate how increasing gain influences the spectral properties of two mode squeezed state. To address this, we numerically solve the theoretical model presented in the previous section.

At low gain, the joint spectral amplitude (JSA) can be found analytically by perturbatively expanding Eq.~(\ref{eq:coupledeq}). The full derivation can be found in Appendix~\ref{app:low_gain_jsa_solution}. It is given by the product of the pump envelope and the phasematching function:
\begin{equation}
	J(\omega_s, \omega_i) = \alpha(\omega_s + \omega_i)\phi(\Delta k(\omega_s, \omega_i)),
	\label{eq:low_gain_jsa}
\end{equation}
where $\phi(\Delta k(\omega_s, \omega_i))$ is the phasematching function, $\alpha(\omega_s + \omega_i)$ is the pump envelope and $\Delta k(\omega_s, \omega_i)$ is the phase mismatch. The shape and orientation of the JSA in the $(\omega_s, \omega_i)$ space is strongly influenced by the phasematching function, which is in turn determined by the group velocities of the interacting fields. Although our formalism accounts for strong dispersion, for the present analysis, we will focus on a simplified case in which the interacting fields are considered to be sufficiently narrow-band to allow a linear expansion of the dispersion relations of the involved modes. This will allow a clear geometric interpretation of the frequency correlations of the output state, and enable us to focus on the underlying physical insights. Under the aforementioned assumption, it is justified to expand the phase mismatch $\Delta k(\omega_s, \omega_i)$ around the central, phasematched signal and idler frequencies~\cite{Keller}. Thus, we obtain
\begin{equation}
	\begin{aligned}
		\Delta k(\omega_s, \omega_i) \approx 
		&\left( \frac{1}{v^{s}_{g}} - \frac{1}{v^{p}_{g}} \right)(\omega_s - \omega_{s0}) \\
		&+ \left( \frac{1}{v^{i}_{g}} - \frac{1}{v^{p}_{g}} \right)(\omega_i - \omega_{i0}),
	\end{aligned}
\end{equation}
where $v^{p}_{g}$, $v^{s}_{g}$, and $v^{i}_{g}$ are the group velocities for the pump, signal, and idler modes, respectively.

To systematically study the gain-dependent effects on the spectral properties, we investigate different group-velocity combinations in the WG, which result in varied phasematching slopes. We characterize these slopes in terms of an angle $\theta$, defined with respect to the $\omega_s$-axis in the $(\omega_s, \omega_i)$ plane \cite{Christ2009}. The angle $\theta$ is expressed in terms of the group velocities of the pump, signal, and idler modes in the PDC process as:  
\begin{equation}
	\theta = \arctan\!\left[ -\left( \frac{v^{s}_{g} - v^{p}_{g}}{\,v^{i}_{g} - v^{p}_{g}} \right)\frac{v^{i}_{g}}{v^{s}_{g}} \right].
	\label{eq:theta}
\end{equation}

\begin{figure}[!htbp]
	\centering
	\includegraphics[width=0.45\textwidth, keepaspectratio]{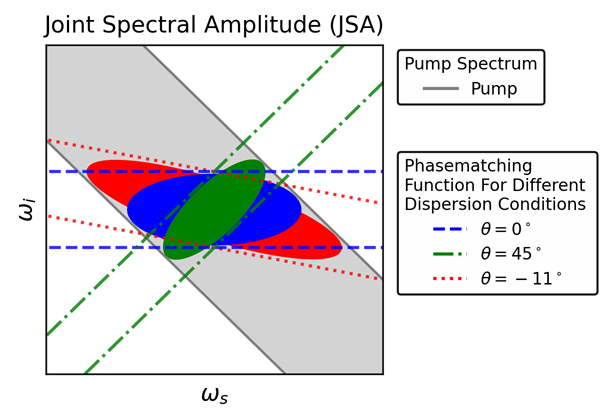}
	\caption{Illustration of  the three dispersion conditions under consideration, characterized by their respective angles ($\theta$) with respect to the frequency ($\omega_s$) axis. These dispersion conditions determine the phasematching characteristics of the system and thereby the shape and orientation of the joint spectral amplitude (JSA) at low gain
	Eq.~(\ref{eq:low_gain_jsa}). Note that these schematics are not exact numerical JSAs but serve as a visual intuition to show the initial states that form the basis for our simulations.}
	\label{fig:jsas}
\end{figure}
We consider three dispersion conditions for a fixed pump bandwidth, as illustrated in
Fig.~\ref{fig:jsas}. The figure schematically depicts the low-gain joint spectral amplitudes
(JSAs) corresponding to phasematching angles $\theta = 0^\circ$, $45^\circ$, and $-11^\circ$.
At low gain, these correspond to different types of spectral correlations: one case yields a nearly uncorrelated spectrum, another produces a strongly correlated spectrum, and the third represents an intermediate situation. Specifically, the three cases correspond to distinct group-velocity relations:
\begin{itemize}
	\item For $\theta = 0^\circ$, the pump and signal modes have the same group velocity, while the idler mode can, in general, be faster or slower than both. In our case, the idler mode group velocity is higher than that of the pump and signal modes.
	\item For $\theta = 45^\circ$, the pump mode group velocity lies midway between those of the signal and idler modes.
	\item For $\theta = -11^\circ$, the idler mode has a higher group velocity than the signal mode, and the signal mode has a higher group velocity than the pump mode.
\end{itemize}

To study the gain-dependent effects of different dispersion conditions, we perform numerical
simulations. Since our primary objective is to investigate the gain-dependent behavior of the
spectrum, we introduce certain simplifications. In particular, all linear optical properties
of the waveguide are assumed to be identical across the different dispersion configurations,
except for the group velocity of the signal mode, which is varied to modify the relative
group-velocity ordering. In addition, the nonlinear overlap integral defined in
Eq.~(\ref{eq:M}) is kept fixed at a realistic value of order
$10^{-6}\,\mathrm{(1/V)}$ for all cases considered.

To systematize the simulations, we ensure that the phasematching bandwidth remains nearly identical across all dispersion conditions by adjusting the WG length accordingly, as it is a key parameter governing the phasematching bandwidth. Similarly, the spectral bandwidth of the pump is kept constant across different dispersion scenarios. However, for each case, we consider both a narrowband $\left(\tfrac{\sigma_p}{\omega_{p_0}} = 0.0015,\; \sigma_t = 274.10\,\mathrm{fs}\right)$
and a broadband $\left(\tfrac{\sigma_p}{\omega_{p_0}} = 0.003,\; \sigma_t = 137.05\,\mathrm{fs}\right)$
pump, where $\omega_{p_0}=2.43\times10^{15}\,\mathrm{1/s}$, in order to also investigate how the pump bandwidth influences the gain-dependent dynamics of PDC spectrum. Table~\ref{tab:param_table} describes the parameters used in the simulations. 
\begin{table}[!htbp]
	\renewcommand{\arraystretch}{1.8} 
	\centering
	\begin{tabular}{|c|c|c|c|c|}
		\hline
		\shortstack{ Dispersion \\ Condition \\ ($\theta$)} & 
		\shortstack{ Waveguide \\ Length \\ $L$ (mm)} & 
		\shortstack{ Group \\ velocity \\ pump \\ $v^{p}_{g}$ (m/s)} & 
		\shortstack{ Group \\ velocity \\ signal \\ $v^{s}_{g}$ (m/s)} &  
		\shortstack{ Group \\ velocity \\ idler \\ $v^{i}_{g}$ (m/s)} \\
		\hline
		$0^\circ$ & 2.5 & \large $\frac{c}{2.168}$ &\large $\frac{c}{2.168}$ & \large $\frac{c}{1.909}$ \\ \hline
		$45^\circ$ & 1.66 & \large $\frac{c}{2.168}$ & \large $\frac{c}{2.426}$ & \large $\frac{c}{1.909}$ \\ \hline
		$-11^\circ$ &  2.5 & \large $\frac{c}{2.168}$ & \large $\frac{c}{2.118}$ & \large $\frac{c}{1.909}$ \\ \hline
	\end{tabular}
	\caption{Group velocities and parameters for different dispersion conditions.}
	\label{tab:param_table}
\end{table}
A further point we investigate is how different poling profiles affect the gain-dependent evolution of the spectrum. Poling allows one to tailor the nonlinear response of a WG, and different patterns lead to distinct phasematching functions \cite{HUM2007180}. In our study, we consider two cases: unapodized WGs, which yield a sinc-type phasematching function \cite{Ayhan2025}, and apodized WGs, engineered to produce a smoother, Gaussian-like phasematching profile \cite{Huang2006}. A detailed discussion of these phasematching types is provided in Appendix~\ref{app:phasematching}. The influence of these two phasematching profiles on the spectral correlations of the generated two-mode squeezed states is examined in the following subsections.

To obtain the numerical results, we solve the coupled-mode equations [Eq.~\eqref{eq:coupledeq}] for the signal, and idler modes for each phasematching type. This yields the evolution of the complex functions introduced in the theory section, which fully encode the field properties of the system. From these functions, we construct the second-order moment of the output state as defined in Eq.~\eqref{eq:spectral_corr}. For the Schmidt-mode analysis, we perform a singular value decomposition (SVD) of this second-order moment, which provides the set of squeezing parameters. From these squeezing parameters, we compute the purity [Eq.~\eqref{eq:purity}], the mode contributions [Eq.~\eqref{eq:p_def}], and the gain [Eq.~\eqref{eq:GdB_def}] used throughout this work.

\subsection{Unapodized Case}

\begin{figure*}[!htbp]
	\centering
	\includegraphics[width=\textwidth,keepaspectratio]{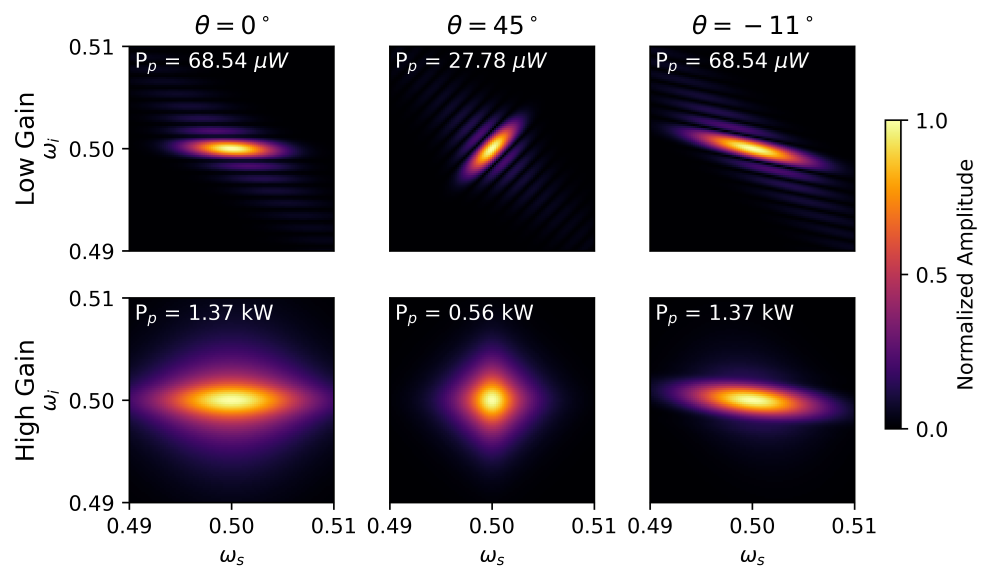} 
	\caption{\label{fig:wide}This figure presents the absolute value of the second moment of the two mode squeezed state [Eq.~(\ref{eq:spectral_corr_schmidt})], which characterizes correlations between the signal and idler modes, for three dispersion conditions defined by the angles $\theta=0^\circ$, $45^\circ$, and $-11^\circ$. For each condition, two gain regimes are shown: low gain (peak powers $\mathrm{P}_p=68.54\,\mu\mathrm{W}$, and $\mathrm{P}_p=27.78\,\mu\mathrm{W}$) and high gain (peak powers $\mathrm{P}_p=1.37\,\mathrm{kW}$, and $\mathrm{P}_p=0.56\,\mathrm{kW}$). All panels are normalized to their respective maxima.}
	\label{fig:sinc jsas} 
\end{figure*}
Fig.~\ref{fig:sinc jsas} shows the simulated second moment of the squeezed state for an unapodized WG under three dispersion conditions, $\theta=0^\circ$, $45^\circ$, and $-11^\circ$, for a broadband pump. The top and bottom rows correspond to the low-gain (peak powers $\mathrm{P}_p=68.54\,\mu\mathrm{W}$ and $27.78\,\mu\mathrm{W}$) and high-gain (peak powers $\mathrm{P}_p=1.37\,\mathrm{kW}$ and $0.56\,\mathrm{kW}$) regimes, respectively. At low gain, the sinc-shaped phasematching function produces visible sidelobes in all three cases. Across the three dispersion conditions, the spectral correlations are qualitatively different: nearly uncorrelated at $\theta=0^\circ$, strongly correlated at $\theta=45^\circ$, and anticorrelated at $\theta=-11^\circ$. At high gain, the second moment broadens and the low-gain sidelobes are suppressed in every case, a well-established effect in the high gain regime~\cite{Spasibko2012}.
\begin{figure*}[!htbp]
	\centering
	\begin{minipage}[t]{0.49\textwidth}
		\centering
		\includegraphics[width=\linewidth]{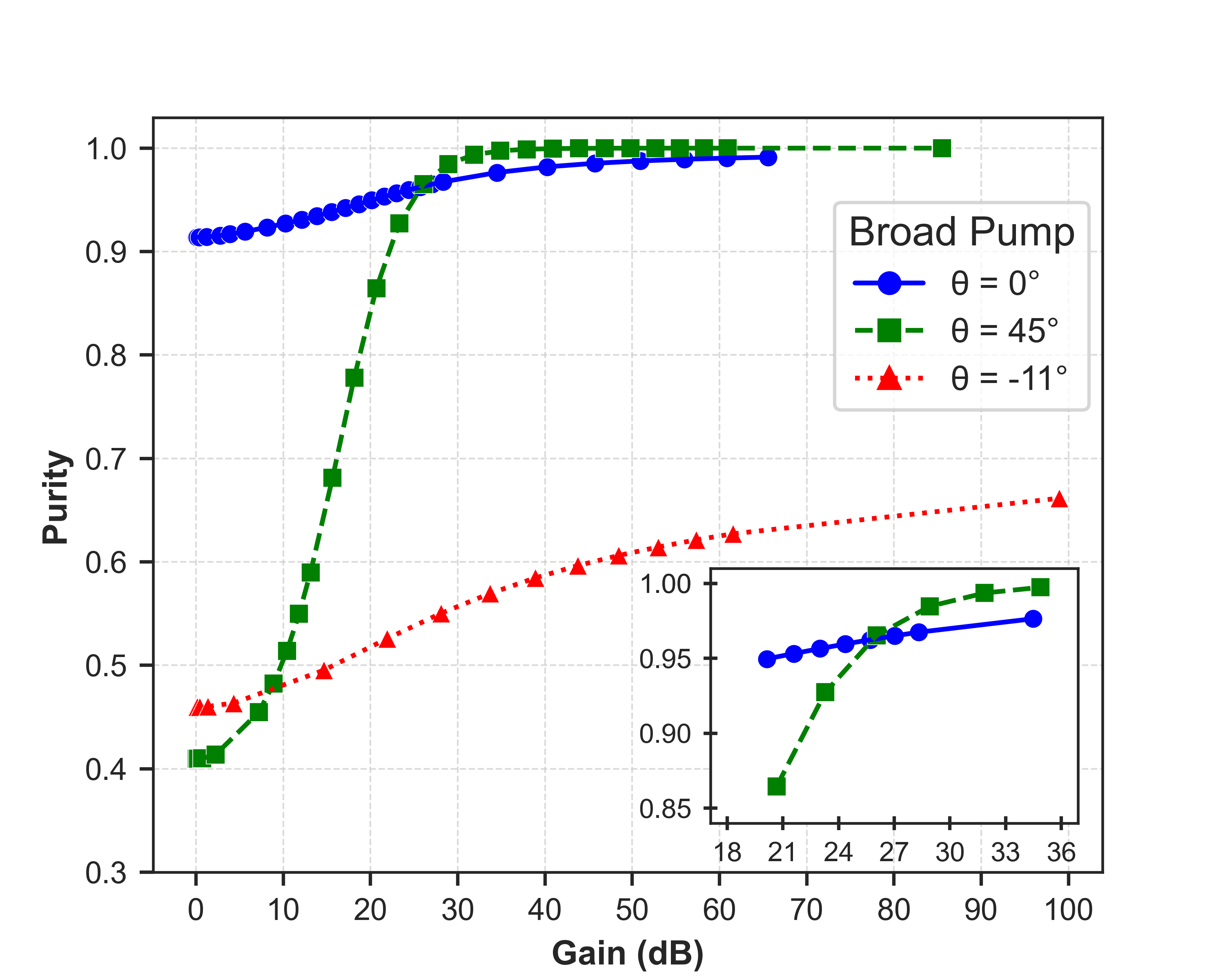}
		\vspace{2pt}\par\textbf{(a)} Broadband Pump Case
	\end{minipage}\hfill
	\begin{minipage}[t]{0.49\textwidth}
		\centering
		\includegraphics[width=\linewidth]{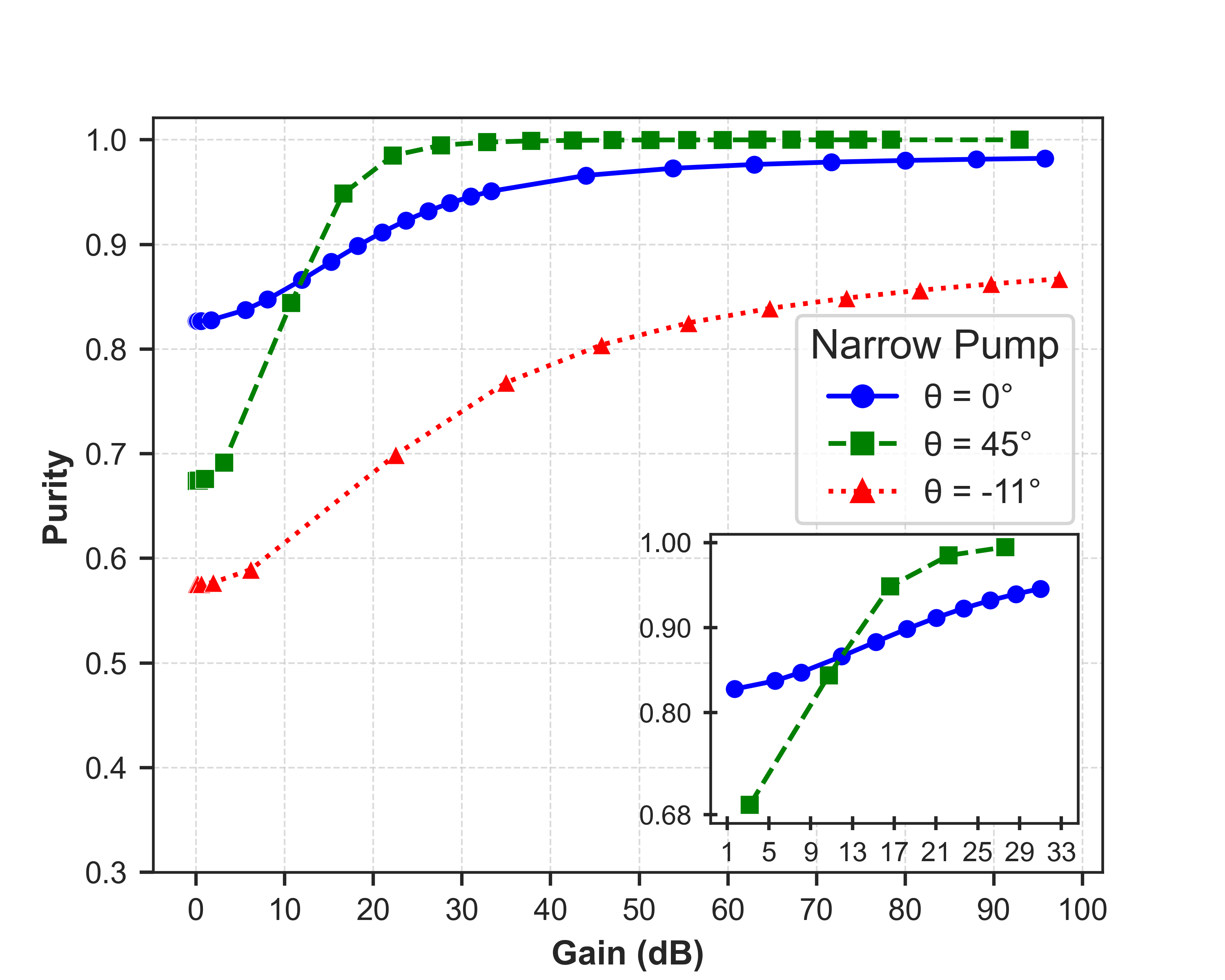}
		\vspace{2pt}\par\textbf{(b)} Narrowband Pump Case
	\end{minipage}
	\caption{This figure shows the relationship between spectral purity and gain for three dispersion conditions defined by angles $\theta = 0^\circ$, $45^\circ$, and $-11^\circ$, under both broadband and narrowband pump conditions. The inset highlights the crossing point of the spectral purity curves for the $\theta = 45^\circ$ and $\theta = 0^\circ$ conditions.
	}
	\label{fig:gain_purity_sinc}
\end{figure*}

Fig.~\ref{fig:gain_purity_sinc} illustrates the relationship between parametric gain and spectral purity for the considered dispersion conditions and pump bandwidths. Fig.~\ref{fig:gain_purity_sinc}(a) corresponds to the broadband pump and shows that, at low gain, the dispersion conditions $\theta=0^\circ$, $45^\circ$, and $-11^\circ$ have purities $(\mathcal{P})$ 0.913 (at $G_\mathrm{dB}=0.038\,\mathrm{dB}$), $0.409$ (at $G_\mathrm{dB}=0.022\,\mathrm{dB}$), and $0.459$ (at $G_\mathrm{dB}=0.042\,\mathrm{dB}$), respectively. As the gain increases, the purity rises in all the cases. However, the three dispersion conditions exhibit distinct gain-dependent dynamics: for $\theta=0^\circ$ the state is nearly uncorrelated at low gain and its purity increases only gradually, for $\theta=45^\circ$ the initially highly correlated state shows a rapid increase in purity with gain, and for $\theta=-11^\circ$ the purity increases at an intermediate rate. At high gain, the purities reach $0.991$ (at $G_\mathrm{dB}=65.57\,\mathrm{dB}$), $0.999$ (at $G_\mathrm{dB}=85.52\,\mathrm{dB}$), and $0.661$ (at $G_\mathrm{dB}=98.95\,\mathrm{dB}$) for $\theta=0^\circ$, $45^\circ$, and $-11^\circ$, respectively.

Notably, although the $\theta=0^\circ$ case starts with substantially higher spectral purity at low gain, the $\theta=45^\circ$ curve overtakes it at intermediate gain [inset of Fig.~\ref{fig:gain_purity_sinc}(a)], with the crossing occurring near $G_\mathrm{dB}\approx 26\,\mathrm{dB}$: $\theta=45^\circ$ attains $\mathcal{P}=0.965$ at $G_\mathrm{dB}=26.04\,\mathrm{dB}$, while $\theta=0^\circ$ yields $\mathcal{P}=0.962$ at $G_\mathrm{dB}=25.70\,\mathrm{dB}$. The initially correlated case also approaches $\mathcal{P}\approx 1$ at much lower gain than $\theta=0^\circ$ case. This demonstrates that a state with strong low-gain correlations can achieve higher purity at high gain than a nearly factorable state. Finally, the $\theta=-11^\circ$ case, where the purity remains well below unity over the tested gain range, demonstrates that there exist dispersion configurations which maintain a fundamentally multimode character, even at extreme values of gain.

In Fig.~\ref{fig:gain_purity_sinc}(b), the narrowband pump case is presented for the same dispersion conditions ($\theta=0^\circ$, $45^\circ$, and $-11^\circ$). Despite the low gain purities differing significantly between the broadband and narrowband pump cases, the different dispersion configurations exhibit a qualitatively similar trend in purity as a function of gain. This suggests that the observed gain-dependent dynamics are largely independent of pump bandwidth.

\begin{figure*}[!htbp]
	\centering
	
	\begin{minipage}[t]{0.34\textwidth}
		\centering
		\includegraphics[width=\linewidth]{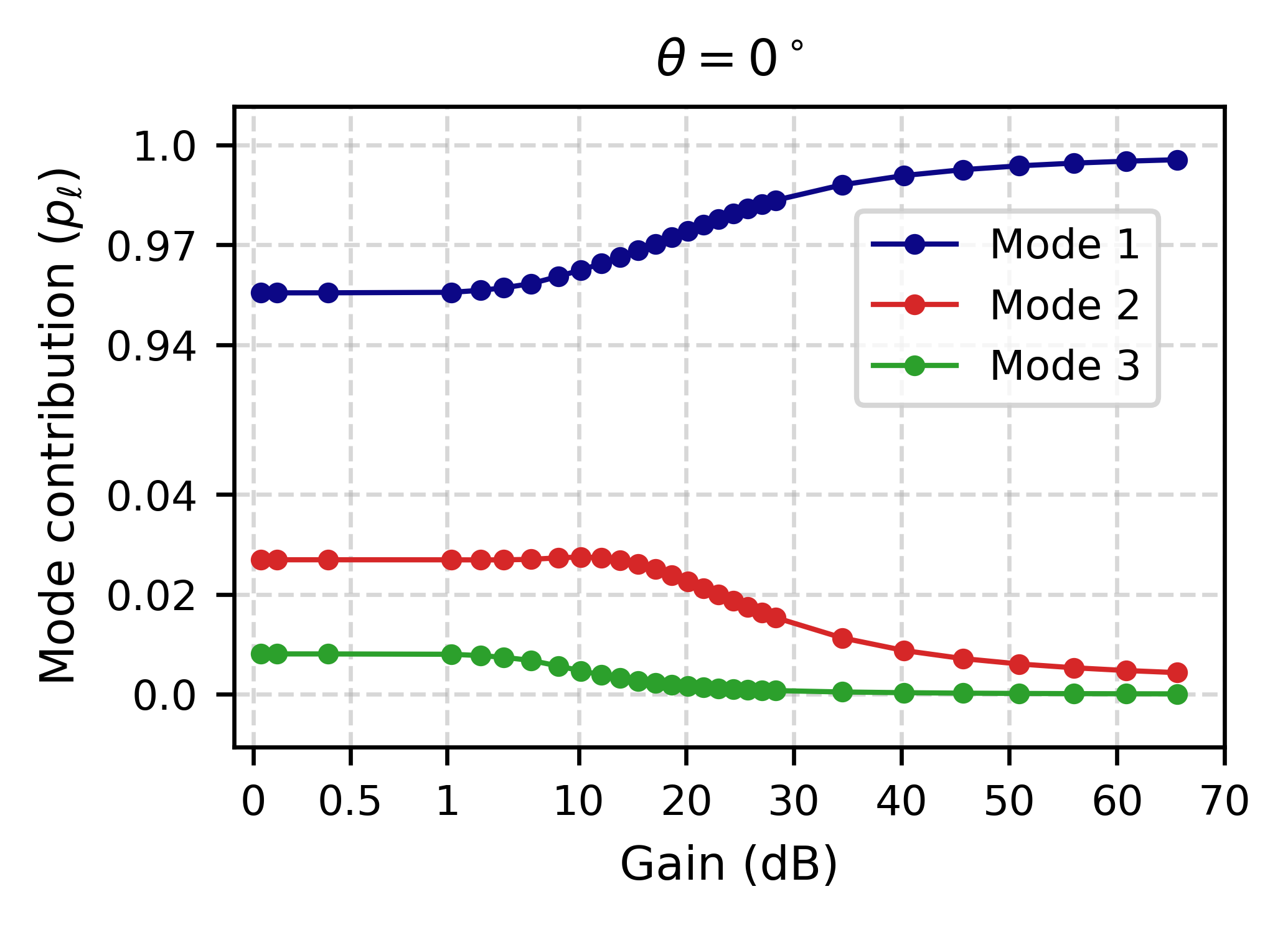}
		\\[1pt]\makebox[\linewidth]{(a)}
	\end{minipage}\hfill
	\begin{minipage}[t]{0.32\textwidth}
		\centering
		\includegraphics[width=\linewidth]{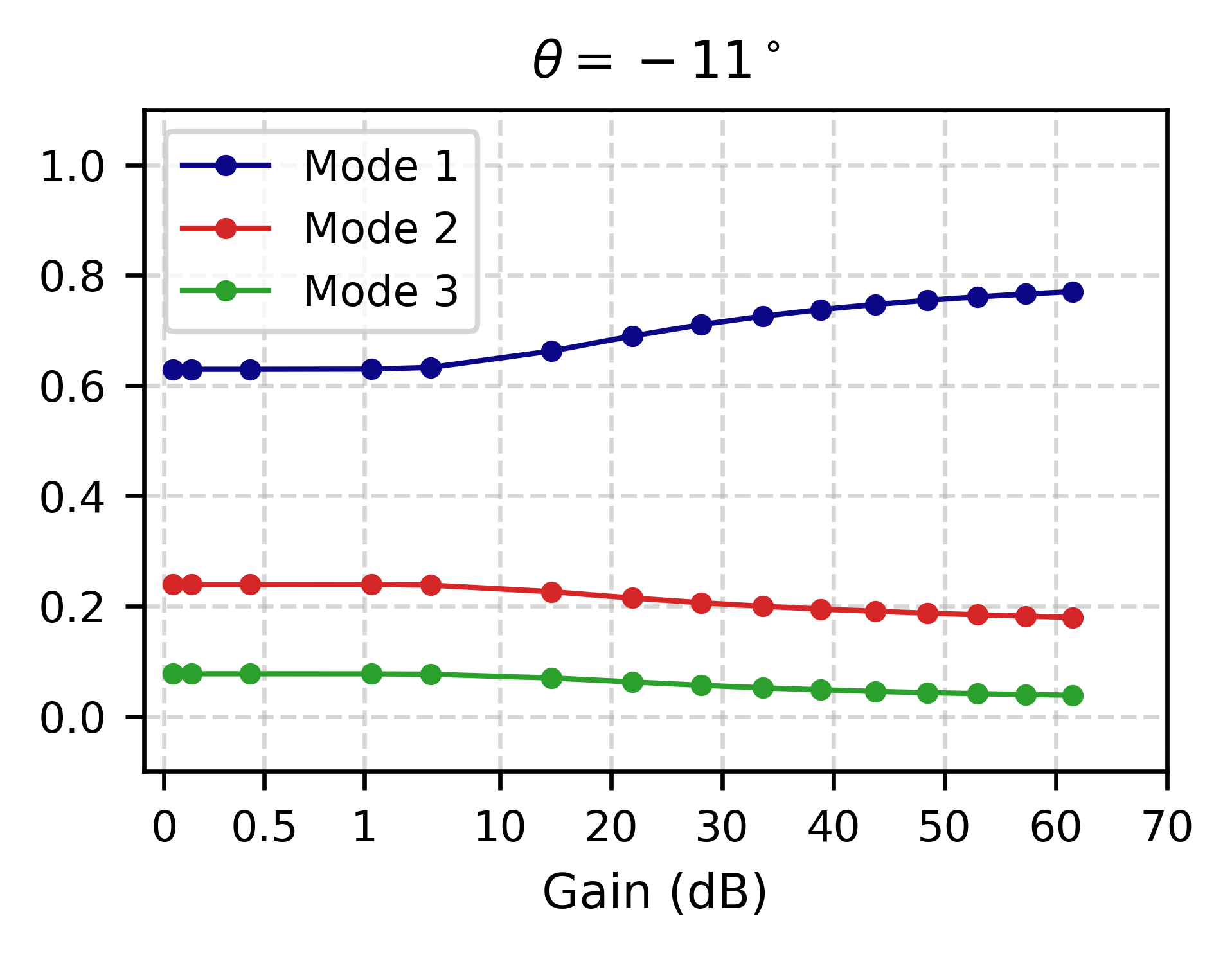}
		\\[1pt]\makebox[\linewidth]{(b)}
	\end{minipage}\hfill
	\begin{minipage}[t]{0.32\textwidth}
		\centering
		\includegraphics[width=\linewidth]{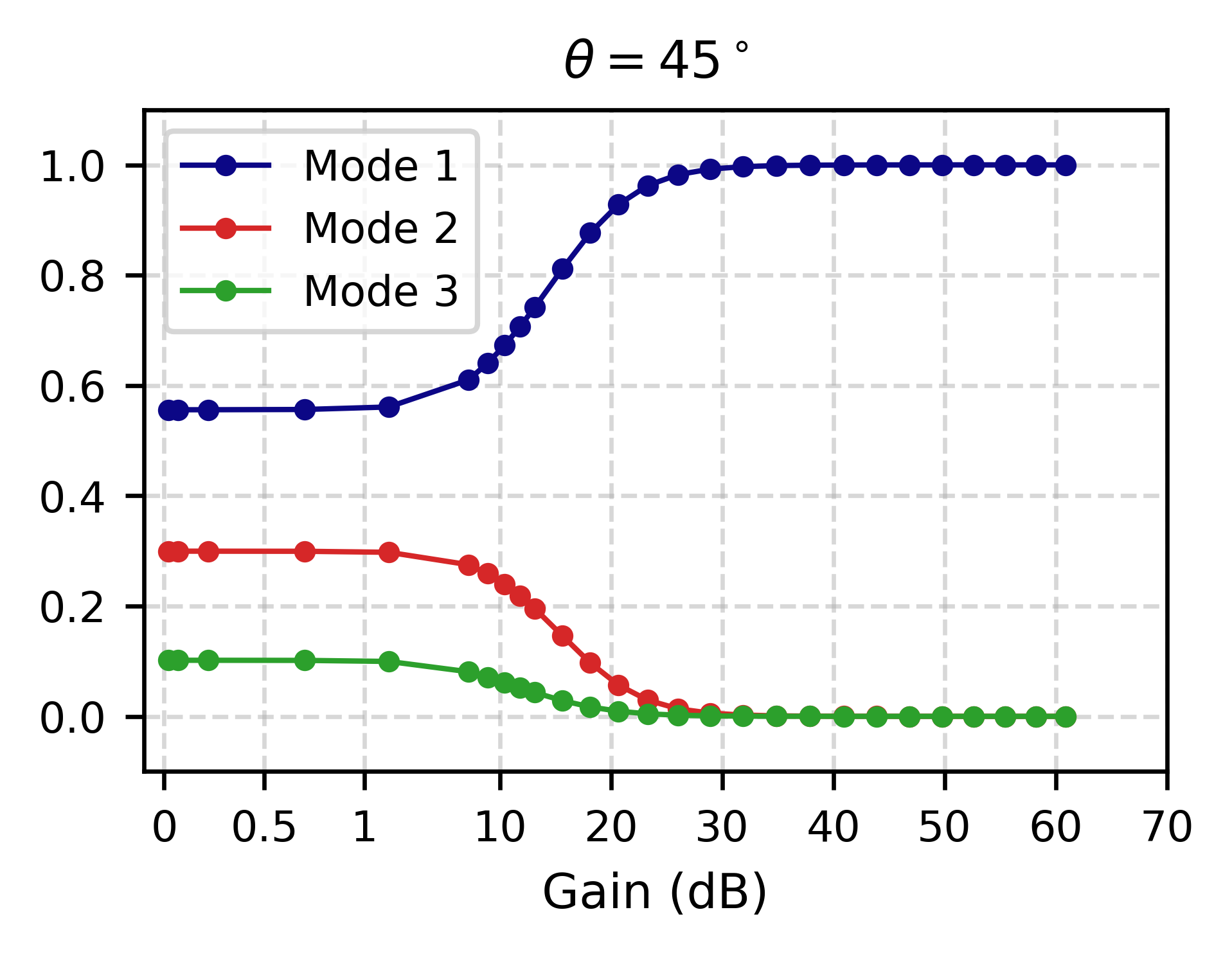}
		\\[1pt]\makebox[\linewidth]{(c)}
	\end{minipage}
	
	\caption{In these figures, for a broadband pump in an unapodized WG (sinc-type phase matching), the mode contribution $p_{\ell}$ of the first three Schmidt modes is plotted versus gain for (a) $\theta=0^\circ$, (b) $\theta=45^\circ$, and (c) $\theta=-11^\circ$.}
	
	\label{fig:sinc_mode_contribution}
\end{figure*}

To gain more insight into the behavior seen in Fig.~\ref{fig:gain_purity_sinc}, we investigate the Schmidt-mode structure of the output spectrum. In particular, we consider the normalized contribution ($p_{\ell}$) of each Schmidt mode to the total spectrum. In this analysis, we focus only on the broadband pump case for conciseness. Fig.~\ref{fig:sinc_mode_contribution} shows the evolution of the normalized contributions of the three lowest-order Schmidt modes ($p_{1,2,3}$) as a function of gain for $\theta = 0^\circ$, $45^\circ$, and $-11^\circ$. In all cases, we see that the the increase in spectral purity observed in Fig.~\ref{fig:gain_purity_sinc} corresponds to a redistribution among the modes, with the lowest-order mode becoming more dominant at increasing gain, while the higher-order contributions decrease. This redistribution is relatively gradual for $\theta = 0^\circ$ and $\theta = -11^\circ$ [Fig.~\ref{fig:sinc_mode_contribution}(a,b)] and, in these cases, is a well-known occurence in high-gain PDC. Namely, frequency components that are already favored by the phasematching function and the pump envelope (i.e., near the central frequencies $\omega_{s0}, \omega_{i0}$) are more populated at lower gain and thus seed further down-conversion events at those frequencies more efficiently. As a result, the lowest-order Schmidt mode (which is centered around those frequencies) is preferentially amplified and its relative weight increases naturally with gain, while contributions from higher-order modes become less prominent~\cite{Sharapova2020,Sharapova2018,Andreas,Christ2011}.

\begin{figure}[!htbp]
	\centering
	\subfloat[\label{fig:squeezing_contributions_sinc_a}]{
		\includegraphics[width=0.45\textwidth]{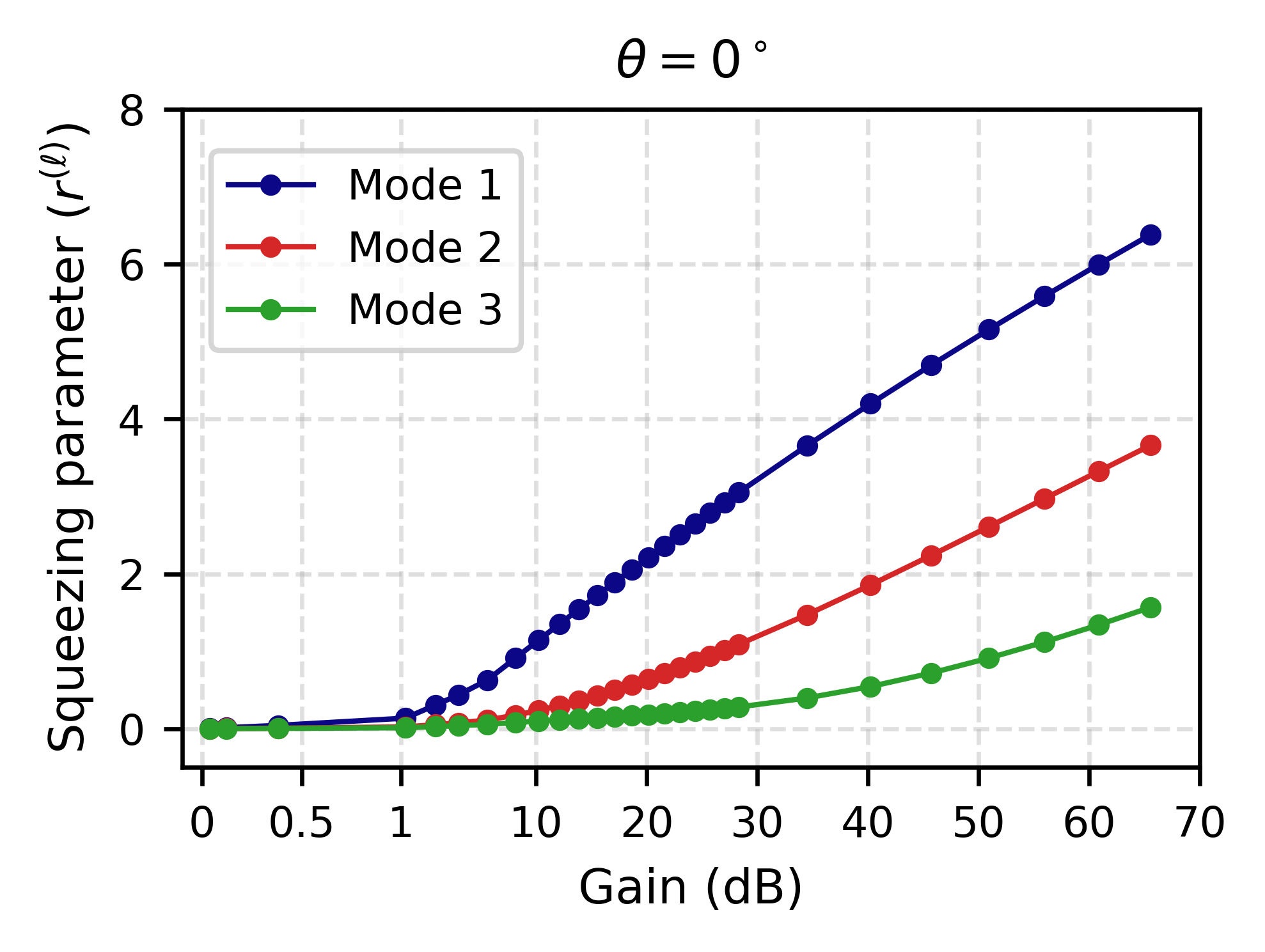}
	}
	\hspace{0.5cm}
	\subfloat[\label{fig:squeezing_contributions_sinc_b}]{
		\includegraphics[width=0.45\textwidth]{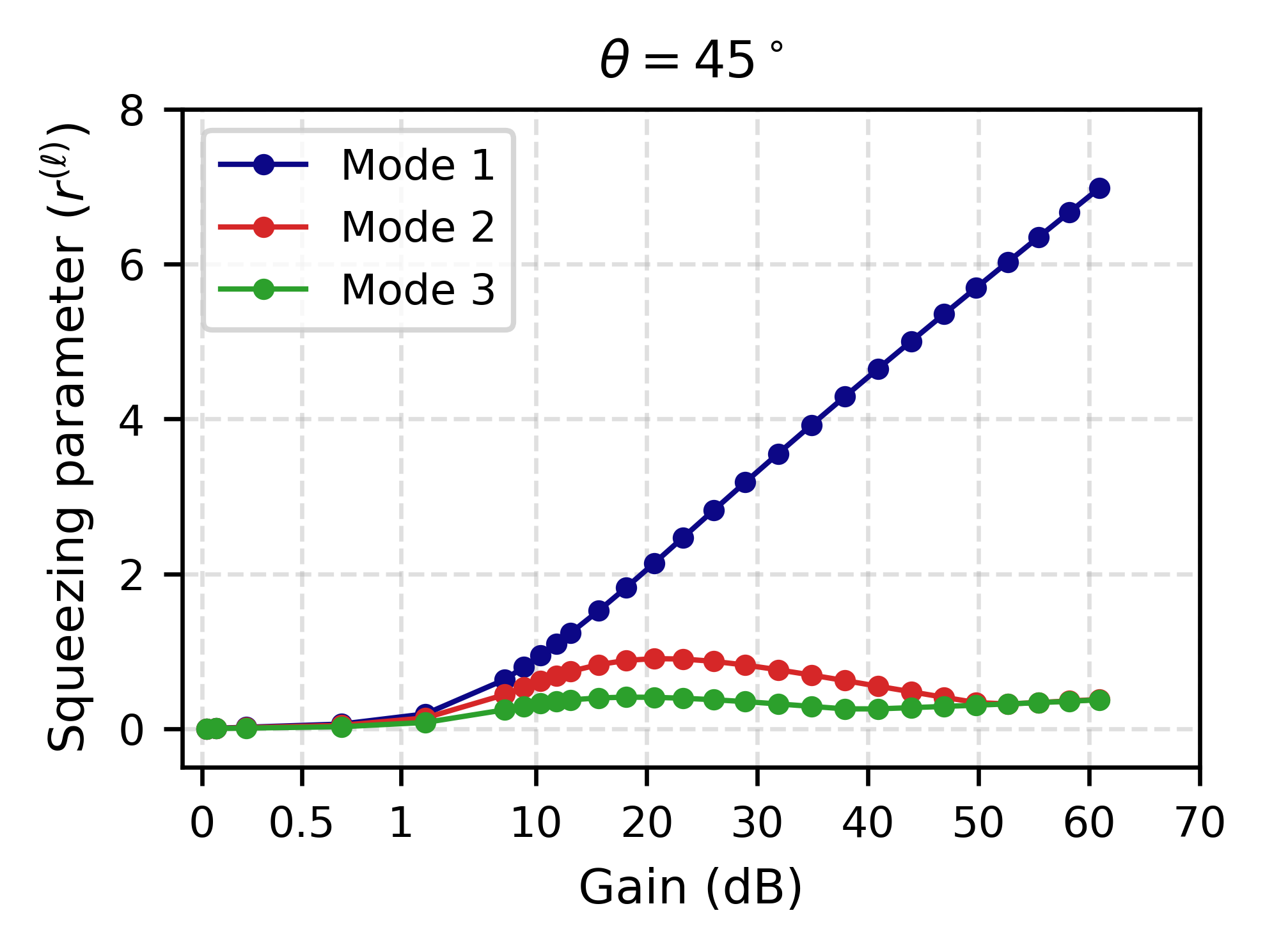}
	}
	\caption{Evolution of the first three squeezing parameters ($r^{(\ell)}$) as a function of gain for (a) $\theta = 0^\circ$ and (b) $\theta = 45^\circ$, shown for the unapodized WG under broadband pumping.}
	\label{fig:squeezing_contributions}	
\end{figure}

For $\theta = 45^\circ$ [Fig.~\ref{fig:sinc_mode_contribution}(c)], the higher-order mode contributions undergo a much sharper decline as gain increases, while the dominant mode $p_1$ rises steeply toward unity. As this redistribution of mode contributions is much faster than the other dispersion configurations, it is reasonable to assume that selective amplification of the lowest-order Schmidt mode mentioned above is here also accompanied by an additional mechanism that further suppresses higher-order Schmidt modes. To verify this, in Fig~\ref{fig:squeezing_contributions}, we plot the corresponding squeezing parameters $r^{(\ell)}$ of the three lowest-order Schmidt modes as a function of gain for the $\theta = 45^\circ$ configuration, as well as $\theta = 0^\circ$ for comparison. We first consider $\theta = 0^\circ$ as a baseline, shown in Fig~\ref{fig:squeezing_contributions}(a). We see that the squeezing factors for all of the Schmidt modes increase monotonically with gain, as predicted by previous works~\cite{Sharapova2020,Sharapova2018}. In contrast, Fig~\ref{fig:squeezing_contributions}(b) shows that the squeezing factors for the higher-order Schmidt modes in the $\theta = 45^\circ$ case reach a maximum value and then \textit{decreases} with further gain. As this effect occurs within the same gain range as the rapid transfer of mode contributions in Fig.~\ref{fig:sinc_mode_contribution}(c), we conclude that this reversed behavior of higher-order squeezing factors is the additional cause of the rapid purification of the spectrum in the $\theta = 45^\circ$ dispersion configuration.

To our knowledge, this kind of behavior has not been reported before in works on the spectral properties of high-gain, single-pass PDC in WGs. The fact that this effect only occurs in the dispersion configuration $\theta = 45^\circ$, and appears independent of the pump bandwidth, suggest that the particular group-velocity configuration between the pump, signal and idler fields is its main cause. However, before attempting to provide a physical interpretation of this effect, we will investigate whether it persists when the WG is apodized.

\subsection{Apodized Case}
\begin{figure}[!htbp]
	\centering
	\includegraphics[width=0.45\textwidth, keepaspectratio]{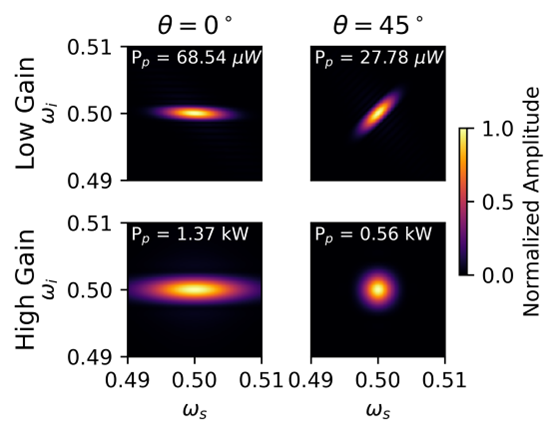}
	\caption{This figure shows the absolute value of the second moment of the squeezed state [Eq.~(\ref{eq:spectral_corr_schmidt})] for Gaussian-type phase matching under two dispersion settings ($\theta=0^\circ$ and $45^\circ$). For each setting, results are shown at two gain levels: low gain with peak pump powers $\mathrm{P}_p=68.54\,\mu\mathrm{W}$ (for $\theta=0^\circ$) and $\mathrm{P}_p=27.78\,\mu\mathrm{W}$ (for $\theta=45^\circ$), and high gain with $\mathrm{P}_p=1.37\,\mathrm{kW}$ (for $\theta=0^\circ$) and $\mathrm{P}_p=0.56\,\mathrm{kW}$ (for $\theta=45^\circ$). All panels are normalized to their respective maxima.}
	\label{fig:gaussian jsas}
\end{figure}

We now simulate the output spectrum of the apodized WG, where the phasematching function is now Gaussian-shaped. Fig.~\ref{fig:gaussian jsas} shows the second moment of this state under two dispersion conditions, $\theta = 0^\circ$ and $45^\circ$, for a broadband pump. Here we omit the $\theta = -11^\circ$ configuration and focus on conditions that approach single-mode behavior at high gain. The top and bottom rows correspond to the low-gain ($\mathrm{P}_p = 68.54\,\mu\mathrm{W}$ and $27.78\,\mu\mathrm{W}$) and high-gain ($\mathrm{P}_p = 1.37\,\mathrm{kW}$ and $0.56\,\mathrm{kW}$) regimes, respectively. At low gain, the Gaussian phasematching profile suppresses the side lobes observed in the unapodized case, as evident in Fig.~\ref{fig:gaussian jsas} (top row). At high gain, the second moment broadens again, consistent with the behavior observed in the unapodized scenario.

\begin{figure}[!htbp]
	\centering
	\includegraphics[width=0.45\textwidth, keepaspectratio]
	{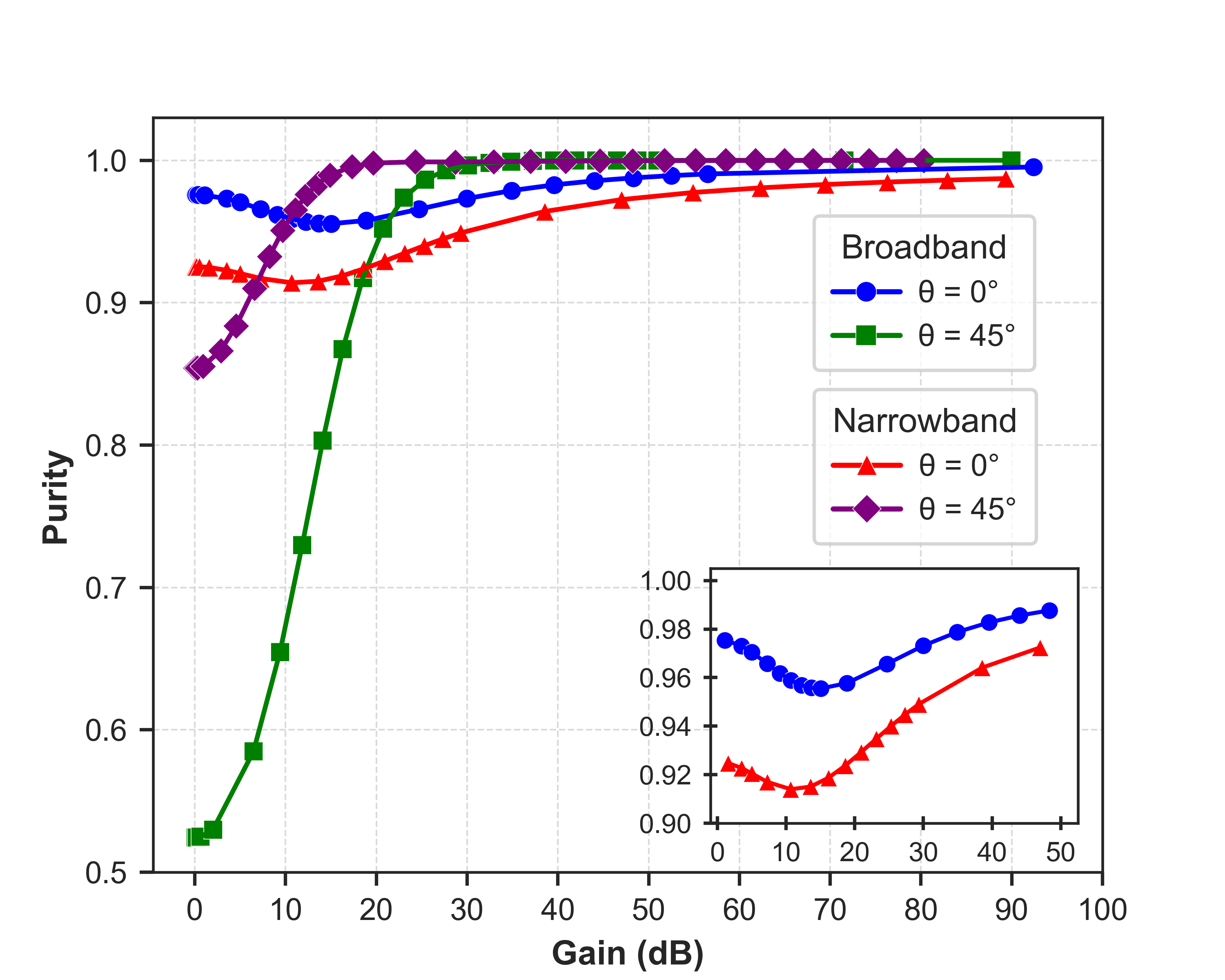}
	\caption{This figure illustrates the relationship between gain and spectral purity for Gaussian phasematching under two dispersion conditions ($\theta = 0^\circ$ and $\theta = 45^\circ$) for both broadband and narrowband pump scenarios. The inset shows the nonmonotonic purity behavior for the $\theta = 0^\circ$ condition.
	}
	\label{fig:gain_purity_gaussian}
\end{figure}

We again quantitatively examine the spectral correlations of the two-mode squeezed state as a function of gain by analyzing its spectral purity. Fig.~\ref{fig:gain_purity_gaussian} shows the gain dependence of spectral purity for Gaussian-type phasematching under both broadband and narrowband pump conditions. In the broadband pump case, the low-gain purities ($\mathcal{P}$) are $0.975$ (at $G_\mathrm{dB}=0.034\,\mathrm{dB}$) for $\theta=0^\circ$ and $0.524$ (at $G_\mathrm{dB}=0.019\,\mathrm{dB}$) for $\theta=45^\circ$. At high gain, although both dispersion conditions eventually reach very high purity, the $\theta=45^\circ$ configuration achieves $\mathcal{P}=0.999$ already at $G_\mathrm{dB}=34.85\,\mathrm{dB}$, whereas the $\theta=0^\circ$ case requires a substantially higher gain ($G_\mathrm{dB}=92.38\,\mathrm{dB}$) to reach $\mathcal{P}=0.995$. For the narrowband pump case, we observe qualitatively similar purity behavior. Thus, the \textit{overall} trend is consistent with the unapodized case: purity improves at high gain, but we again see that $\theta=45^\circ$ reaches higher purity values more rapidly than $\theta=0^\circ$. This similarity suggests that this particular trend is also independent of the exact form of the phasematching function, in addition to the pump bandwidth.

However, unlike the unapodized case, the apodized $\theta=0^\circ$ case shows a distinct nonmonotonic behavior. Although it is nearly pure at low gain it exhibits a clear dip at moderate gain before again recovering at higher gain (see Fig.~\ref{fig:gain_purity_gaussian}, inset). The monotonic decrease of the purity has been discussed in
Refs.~\cite{Houde2023,Thekkadath,Quesada2015}, where it was found to be due to additional phase correlations introduced in the joint spectrum as a consequence of seeding effects. Our results extend this picture by showing that, as gain is further increased, the purity rises again.
\begin{figure}[!htbp]
	\centering
	\subfloat[\label{fig:mode_contributions_gaussian_a}]{
		\includegraphics[width=0.45\textwidth]{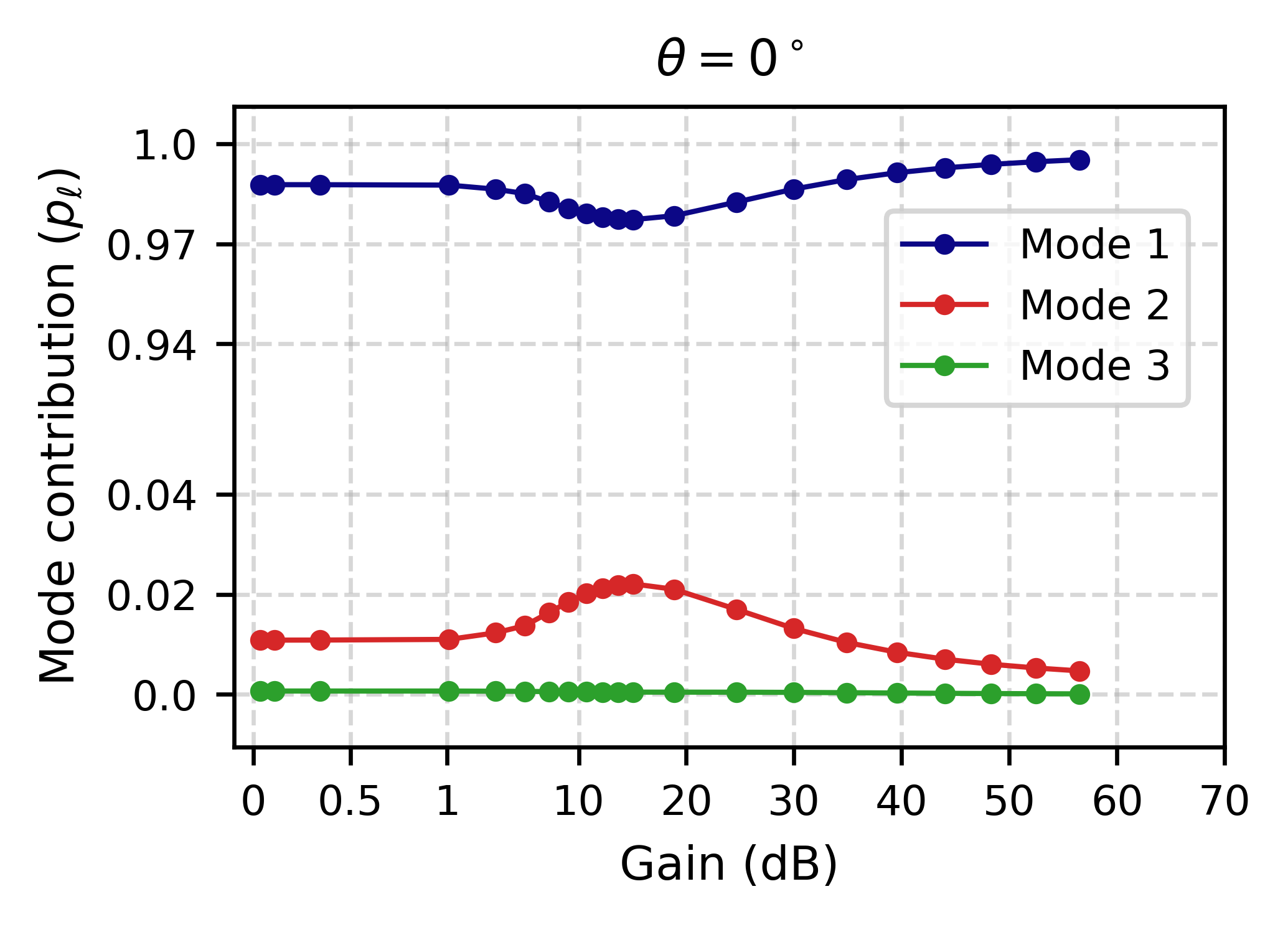}
	}
	\hspace{0.5cm}
	\subfloat[\label{fig:mode_contributions_gaussian_b}]{
		\includegraphics[width=0.45\textwidth]{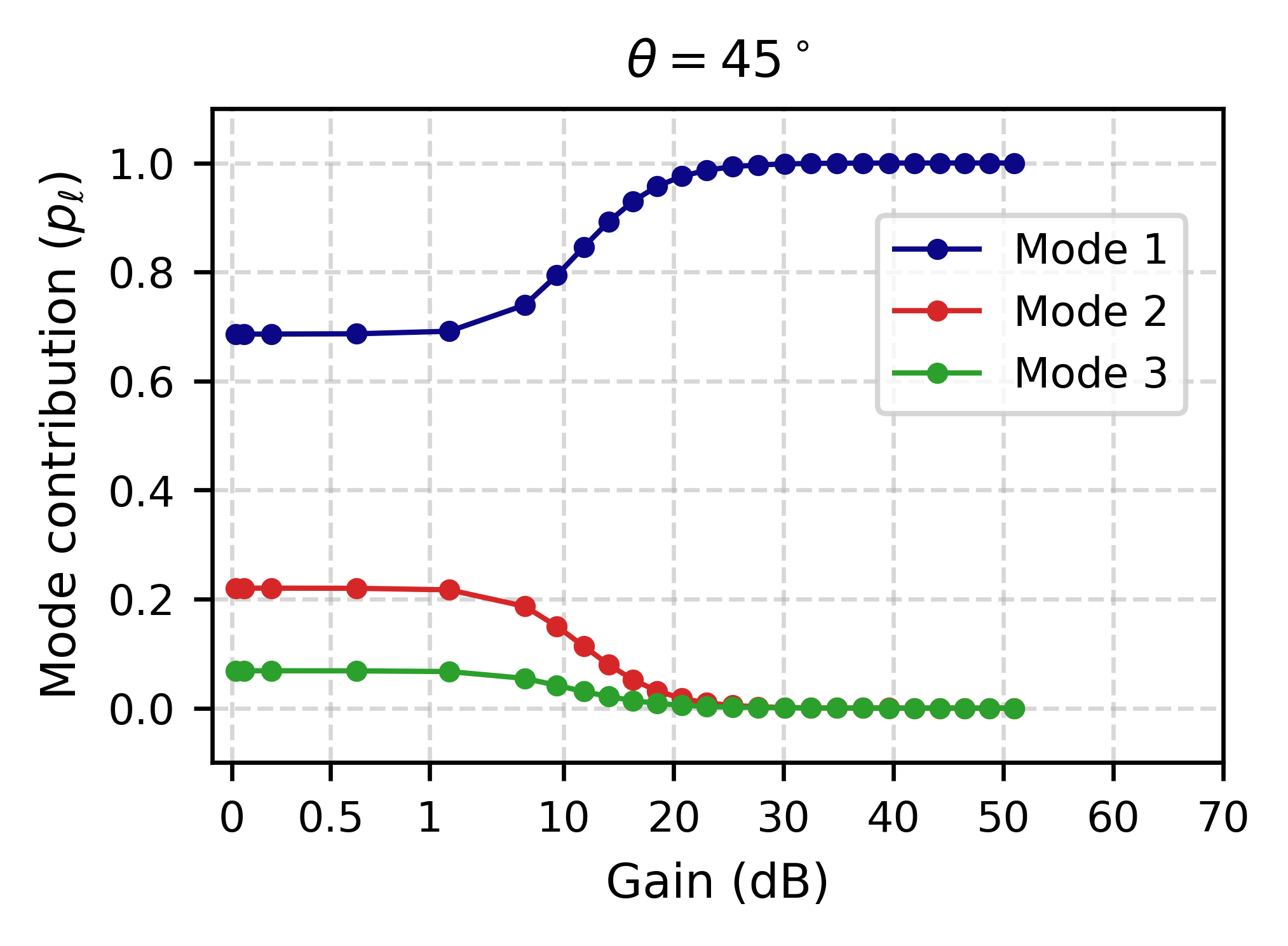}
	}
	\caption{Mode contribution $p_{\ell}$ of the first three Schmidt modes, plotted versus gain for a broadband pump in an apodized (Gaussian-type phase matching) WG: (a) $\theta=0^\circ$, (b) $\theta=45^\circ$.}
	\label{fig:mode_contribution_gaussian}
\end{figure}

To analyse the above behaviors in more detail, we again investigate the gain-dependent Schmidt mode structure of the output spectra and also only focus on the broadband pump for conciseness. In Fig.~\ref{fig:mode_contribution_gaussian}, the contributions of the first three Schmidt modes ($p_k$) are shown for the apodized configuration under two dispersion conditions, $\theta = 0^\circ$ and $\theta = 45^\circ$. At $\theta = 0^\circ$, the dominant mode contribution $p_1$ exhibits a shallow dip at intermediate gain ($G_{\mathrm{dB}} \approx 15\,\mathrm{dB}$) before recovering toward unity. This mirrors the nonmonotonic trend observed in Fig.~\ref{fig:gain_purity_gaussian} and the temporary reduction in purity indeed corresponds to a redistribution of weight into higher-order modes. As gain is further increased, the lowest-order Schmidt mode is again more efficiently amplified, thereby suppressing these gain-induced correlations.

\begin{figure}[!htbp]
	\centering
	\includegraphics[width=0.95\linewidth, keepaspectratio]{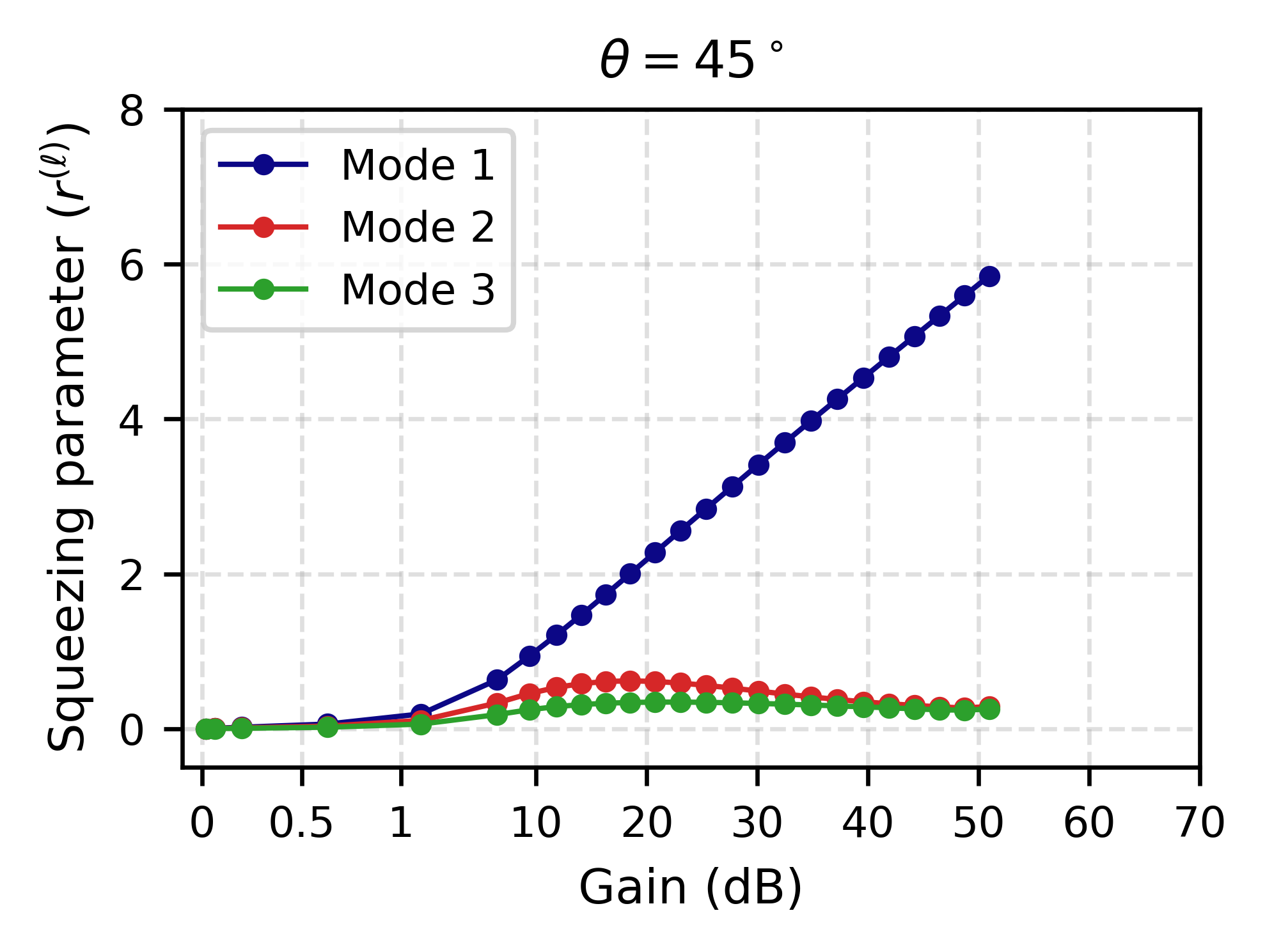}
	\caption{Evolution of the first three squeezing parameters ($r^{(\ell)}$) as a function of gain for the $\theta = 45^\circ$ case in the apodized WG. } 
	\label{fig:squeezing_contributions_gauss}	
\end{figure}

Now we turn our attention to the $\theta = 45^\circ$ case. We see that it follows the same trend as in the unapodized configuration: $p_1$ rapidly increases with gain, while higher-order contributions ($p_2$ and $p_3$) rapidly diminish. To further confirm that this is the same effect as observed in the unapodized WG, we again look at the gain dependence of the squeezing factors $r_k$ for the $\theta = 45^\circ$ configuration. This is shown in Fig.~\ref{fig:squeezing_contributions_gauss} and we indeed see that the higher-order Schmidt modes are being additionally suppressed at high gain. The occurrence of this suppression effect for $\theta = 45^\circ$ also in the apodized WG reaffirms that it is primarily determined by the underlying group-velocity configuration. In the following, we present a physical explanation that accounts for this behavior across both phasematching types.

\subsection{Group-velocity-dependent purification mechanism}\label{subsec:gain_dynamics}

As already indicated, the observed additional suppression of higher-order Schmidt modes at high gain occurs in the $\theta = 45^\circ$ configuration, regardless of pump bandwidth and the type of phasematching. As the phasematching angle $\theta$ is primarily dictated by the relative group velocities of the interacting modes [see Eq.~\eqref{eq:theta}], it is reasonable to assume that such behavior is caused by high-gain seeding effects  amplifying and modulating the effects of the temporal walk-off between pump, signal, and idler \cite{Grice2001,mosley2008heralded,ansari2018tailoring}.

When a pump pulse is sent to the nonlinear WG, over each small longitudinal “slice’’ of the waveguide, the three fields can interact and the contribution of each slice to the output joint spectrum includes a walk-off–induced phase \cite{Grice2001,mosley2008heralded,ansari2018tailoring,Thekkadath,Quesada2015}. At low gain, all slices contribute with approximately equal weight, and the interference of these uniformly weighted contributions produces the familiar low-gain phasematching structure \cite{Grice2001,mosley2008heralded,ansari2018tailoring,uren2006}.

At high gain, however, the situation changes as seeding effects take hold \cite{Quesada2015,Quesada,Andreas, Houde2023,Helt}. Photon pairs generated earlier in the WG propagate for a longer distance before participating in further down-conversion and therefore accumulate more walk-off phase. Furthermore, each subsequent stimulated event inherits the phase of the field that seeded it, so the phase accumulated early in the WG is carried forward and amplified. On the other hand, photons generated later accumulate less phase but are more numerous, as high-gain PDC strongly favours pair creation near the WG output \cite{Thekkadath,uren2006}. As a result, the effective weighting of the longitudinal slices becomes highly non-uniform, and the output joint spectrum, along with the spectral structure of the individual Schmidt modes, changes with gain~\cite{Sharapova2018,Sharapova2020,Thekkadath,uren2006}.

In the $\theta = 45^\circ$ configuration, the group velocities satisfy $v_g^{(s)} < v_g^{(p)} < v_g^{(i)}$ and the signal and idler acquire walk-off phases of opposite sign. Thus the stimulated fields they generate downstream contribute to the output with likewise opposing signs and, as gain is increased, more and more such contributions appear. This introduces a new, \textit{gain-dependent interference} mechanism which is most prominent for spectral components that already possess a nonzero phase at the output at low gain. Thus, it may lead such frequencies to potentially interfere destructively as gain is increased. Since higher-order Schmidt modes possess varying spectral phases, it is exactly their spectral components that are preferentially suppressed by this interference. In contrast, the lowest-order Schmidt mode has an almost flat spectral phase \cite{Andreas,Houde2023}, so its amplification remains unaffected by the walk-off-induced interference.

Crucially, this mechanism does not rely on the perfect symmetry of the $\theta = 45^\circ$ case, which requires $v_g^{(p)}$ to lie halfway between $v_g^{(s)}$ and $v_g^{(i)}$, but to the overall group-velocity ordering $v_g^{(s)} < v_g^{(p)} < v_g^{(i)}$. This is sufficient to ensure that that seeding effects generate frequency components with opposite phases with respect to the pump, leading, in turn, to destructive interference of higher-order Schmidt modes. To further verify this claim, we conducted additional simulations for $\theta = 40^\circ, 42^\circ, 47^\circ,$ and $50^\circ$, which confirm that the same high-gain suppression of higher-order Schmidt modes occurs whenever the pump velocity lies between those of signal and idler. The results of those simulations are included in the Appendix~\ref{app:other_gv_config}, but they confirm that exact symmetry in the group velocity configuration is not required.

In contrast, for the $\theta = 0^\circ$ configuration, $v_g^{(s)} = v_g^{(p)} < v_g^{(i)}$ and the walk-off phase contributions only come from the idler photons. And in the $\theta = -11^\circ$ configuration, $v_g^{(p)} < v_g^{(s)} < v_g^{(i)}$ the contributions from both photons have the same sign. Under such conditions, the effects of the gain-induced interference are much less prominent and there is no additional suppression of higher-order Schmidt modes, in accordance with our simulation results. These observations further confirm that the distinctive behaviour seen in the $\theta = 45^\circ$ case originates from the pump lying between the signal and idler velocities, which alone enables the opposite-sign walk-off phases required for high-gain, interference-driven purification.

\section{Conclusion}

We have studied the spectral properties of both unapodized and apodized dispersion-engineered WGs for the generation of two-mode squeezed states in the high-gain regime. Our results show that the overall behavior of spectral purity as a function of gain, in both configurations, is primarily dictated by the dispersion conditions of the WG.

In the unapodized configuration, we considered three dispersion conditions ($\theta = 0^\circ$, $45^\circ$, and $-11^\circ$), each exhibiting a distinct gain-dependent evolution of spectral purity. For $\theta = 45^\circ$, the purity rises sharply with gain, whereas for $\theta = 0^\circ$ the increase is more gradual, though it still approaches near unity at high gain. In contrast, the $\theta = -11^\circ$ case remains fundamentally multimode, with purity remaining well below unity. In the apodized configuration, the $\theta = 45^\circ$ case again shows a monotonic increase in purity, while the $\theta = 0^\circ$ configuration exhibits nonmonotonic behavior: the purity initially decreases due to gain-induced phase correlations and later recovers at higher gain. Previous studies such as Ref.~\cite{Thekkadath} reported only a monotonic decrease. Our results extend this understanding of gain-dependent dynamics by showing that purity can recover once the lowest-order Schmidt mode becomes strongly dominant.

A key outcome of our study is the identification of a distinct high-gain suppression mechanism for higher-order Schmidt modes. In the dispersion configuration where the pump group velocity lies between those of the signal and idler ($v_g^{(s)} < v_g^{(p)} < v_g^{(i)}$, corresponding to $\theta = 45^\circ$), the squeezing parameters of higher-order Schmidt modes do not merely saturate but begin to decrease at sufficiently high gain. This leads to a markedly faster purification of the output state compared to other dispersion settings. By combining Schmidt-mode analysis with a group-velocity-based interpretation, we attribute this behavior to a gain-induced interference mechanism driven by the opposite-sign walk-off phases of the signal and idler fields. Additional simulations for nearby dispersion angles ($\theta = 40^\circ, 42^\circ, 47^\circ, 50^\circ$) show that this effect is robust and does not rely on fine tuning or perfect symmetry; it appears whenever the pump group velocity lies between those of the signal and idler. Moreover, the gain-dependent behavior, including the rapid purification in the $\theta = 45^\circ$ configuration, is insensitive to pump bandwidth and to the detailed form of the phasematching function (sinc or Gaussian). This demonstrates that the mechanism is fundamentally dictated by the dispersion relations of the interacting modes, rather than by the pump bandwidth or phasematching function.

From an experimental standpoint, the $\theta = 45^\circ$ configuration is especially attractive because it allows the degree of spectral entanglement to be tuned solely through the parametric gain. As a result, the same WG can function either as a strongly entangled, multimode source or as an almost single-mode source, depending on the desired regime. This gain-based tunability provides a flexible and compact tool for integrated quantum photonics.

In summary, our work highlights the central role of dispersion engineering in shaping high-gain PDC spectra and uncovers a previously unreported mechanism through which specific group-velocity configurations induce accelerated purification at high gain. These insights provide practical design principles for developing squeezed-light sources tailored to the requirements of diverse quantum technologies.

\begin{acknowledgments}
This research is funded by the German Federal Ministry of Research, Technology and Space (BMFTR, projects 13N16108 PhoQuant and 13N16975 LichtBriQ). This research is supported by funding from the Carl-Zeiss-Stiftung (CZS Center QPhoton). S.S. acknowledges funding by the Nexus program of the Carl-Zeiss-Stiftung (project MetaNN).
\end{acknowledgments}

\appendix

\section{Low-Gain Solution of the Joint Spectral Amplitude (JSA)}
\label{app:low_gain_jsa_solution}
In the low-gain regime, we solve the coupled integral Eq.~(\ref{eq:coupledeq}) from the main text perturbatively. We begin by initializing the complex function $A(\omega_s, \omega^{''}_s, t)$ at the initial time $t_0$, and then solve the following time-dependent differential equation to determine the evolution of the complex function $D(\omega^{'}_i, \omega^{''}_s, t)$:

\begin{equation}
\partial_t D(\omega^{'}_i, \omega^{''}_s, t) = -\frac{i}{\hbar} \int_0^{\infty} \dd{\omega_s} \, G_2(\omega_s, \omega^{'}_i, t) \, A^*(\omega_s, \omega^{''}_s, t).
\end{equation}

We impose the initial condition:
\begin{equation}
A(\omega_s, \omega^{''}_s, t_0) = \delta(\omega_s - \omega^{''}_s),
\end{equation}
and substitute this into the evolution equation to obtain:
\begin{align}
\partial_t D(\omega^{'}_i, \omega^{''}_s, t) 
&= -\frac{i}{\hbar} \int_0^{\infty} \dd{\omega_s} \, G_2(\omega_s, \omega^{'}_i, t) \, \delta(\omega_s - \omega^{''}_s) \nonumber \\
&= -\frac{i}{\hbar} \, G_2(\omega^{''}_s, \omega^{'}_i, t).
\end{align}

Integrating both sides with respect to time yields the solution:
\begin{equation}
D(\omega^{'}_i, \omega^{''}_s, t) = -\frac{i}{\hbar} \int \dd{t} \, G_2(\omega^{''}_s, \omega^{'}_i, t).
\end{equation}

We now express $G_2(\omega^{''}_s, \omega^{'}_i, t)$ using system parameters and obtain:
\begin{align}
D(\omega^{'}_i, \omega^{''}_s, t) 
&= -i  \frac{ M\gamma_p}{2 \pi c} 
\sqrt{ \omega^{'}_i \, \omega^{''}_s \, 
n_g^i(\omega^{'}_i) \, n_g^s(\omega^{''}_s) n_g^p(\omega_{p_0})} \nonumber \\
&\quad \times \int_0^{\infty} \dd{\omega_p} \, 
\alpha(\omega_p) \,\phi(\frac{\Delta k(\omega^{''}_s, 
\omega^{'}_i, \omega_p)L}{2}) 
 \\
&\quad \times \int_{t_0}^{t} \dd{t} \, 
\exp\left(-i (\omega_p - \omega^{''}_s - \omega^{'}_i) t \right).
\end{align}

Evaluating the time integral assuming that there is long interaction time which is given as follows:
\begin{equation}
\int_{-\infty}^{\infty} \dd{t} \exp\left(-i (\omega_p - \omega^{''}_s - \omega^{'}_i) t \right) = 2\pi \delta(\omega_p - \omega^{''}_s - \omega^{'}_i),
\end{equation}
which yields the simplified expression:
\begin{align}
D(\omega^{'}_i, \omega^{''}_s) 
&=  -i \frac{M\gamma_p}{c} 
\sqrt{ \omega^{'}_i \, \omega^{''}_s \, 
n_g^i(\omega^{'}_i) \, n_g^s(\omega^{''}_s) n_g^p(\omega_{p_0})} \nonumber \\
&\quad \times \alpha(\omega^{'}_i + \omega^{''}_s) \,\phi\left(\frac{\Delta k(\omega^{''}_s, 
\omega^{'}_i, \omega^{'}_i + \omega^{''}_s)L}{2}\right).
\end{align}

Next, we compute the second-order moment using Eq.~(\ref{eq:spectral_corr}):
\begin{equation}
\begin{split}
\bra{0}\Tilde{a}_{\omega_s}(t)\Tilde{b}_{\omega_i}(t)\ket{0}
&= \int \dd{\omega_{s}^{''}} A(\omega_s, \omega^{''}_s, t) D(\omega_i, \omega^{''}_s, t) \\
&= \int \dd{\omega_{s}^{''}} \delta(\omega_s - \omega^{''}_s) D(\omega_i, \omega^{''}_s) \\
&= D(\omega_i, \omega_s).
\end{split}
\end{equation}

Thus, the joint spectral amplitude (JSA) in the low-gain regime is given by:
\begin{equation}
\begin{split}
J(\omega_s, \omega_i) &= D(\omega_i, \omega_s) \\
&= \zeta \, \alpha(\omega_s + \omega_i) \, \phi\left(\frac{\Delta k(\omega_s, \omega_i)L}{2}  \right),
\label{eq:lowgainjsa}
\end{split}
\end{equation}
where the prefactor $\zeta$ is calculated assumed that the overlap integral and group indexes have small variations with respect to the frequency which is given as:
\begin{align}
\zeta &= -i \frac{\gamma_p M }{c} 
\sqrt{ \omega_i \omega_s \, 
n_g^i(\omega_i) \, n_g^s(\omega_s) n_g^p(\omega_{p_0})},
\end{align}


This expression captures the spectral correlations at low gain.

\section{Generalized Phasematching}
\label{app:phasematching}
The generalized phasematching function in Eq.~(\ref{eq:phasematching_function}) depends on the
longitudinal modulation of the nonlinear coefficient, which we model as
\[
\chi^{(2)}(z)=d\,f(z),\qquad f(z)\in\{+1,-1\}.
\]
In periodically poled waveguides, \( f(z) \) alternates between \(+1\) and \(-1\) with a fixed
poling period \( \Lambda \) and duty cycle \( D\in(0,1) \). This periodic reversal of the nonlinear
coefficient implements quasi-phase matching (QPM) by compensating the accumulated phase mismatch
between the interacting waves.

\medskip
To understand how the choice of poling pattern determines the phasematching function, it is useful
to expand the modulation function \( f(z) \) into a Fourier series~\cite{HUM2007180}:
\[
f(z)=\sum_{m\in\mathbb{Z}} f_m\,e^{i m K z}, \qquad K=\frac{2\pi}{\Lambda},
\]
with Fourier coefficients
\[
\begin{aligned}
	f_m &= \frac{1}{\Lambda}\int_0^\Lambda f(z)\,e^{-i m K z}\,dz \\[6pt]
	&=
	\begin{cases}
		2D-1, & m=0, \\[6pt]
		\displaystyle \frac{2}{m\pi}\, e^{-i m\pi D}\,\sin(m\pi D), & m\neq 0~.
	\end{cases}
\end{aligned}
\]
In practice, periodically poled waveguides are fabricated with a 50\% duty cycle
($D = 1/2$) \cite{Fiorentino2007,Martin_2010}. In this case, the DC component of the nonlinear grating vanishes
($f_0 = 0$), meaning that the average nonlinear coefficient over one period is zero,
and only odd Fourier harmonics remain. The first-order coefficients are
\[
f_{\pm1} = \mp\,\frac{2i}{\pi}, \qquad |f_1| = \frac{2}{\pi}.
\]
Higher-order harmonics ($|m| \ge 3$) oscillate much more rapidly along the
waveguide and therefore contribute only weakly to the phasematching integral, as
they largely average out over the device length. For this reason, experimental
analyses typically retain only the dominant first-order QPM term. Under this
approximation, the spatial modulation of the nonlinear coefficient reduces to
\[
f(z) \approx f_1 e^{iKz} + f_{-1} e^{-iKz}
= \frac{4}{\pi}\,\sin(Kz),
\]
and the effective nonlinear coefficient becomes
\[
d_{\mathrm{eff}} = \frac{2}{\pi}\, d.
\]

\medskip
When this modulation is integrated over the crystal length, the resulting phasematching function is
\[
\phi\!\left(\frac{\Delta k L}{2}\right)
= 
\frac{2}{\pi}\,
\mathrm{sinc}\!\left[\frac{(\Delta k - 2\pi/\Lambda) L}{2}\right]
\exp\!\left[i \frac{(\Delta k - 2\pi/\Lambda)L}{2}\right].
\]
This sinc-type phasematching response (the unapodized case used in the main text) arises from the first-order quasi-phase-matched interaction integrated over a finite crystal length \(L\). Its characteristic side lobes introduce spectral oscillations in the joint spectrum and are one of the dominant sources of spectral correlations in low-gain PDC.

\medskip
In contrast, an \emph{apodized} (aperiodically poled) waveguide intentionally modulates the poling pattern slowly along the propagation direction. By allowing
the poling period \( \Lambda(z) \) or the duty cycle \( D(z) \) to vary gradually with \(z\), the amplitude and phase of the first Fourier harmonic acquire a smooth
spatial envelope \cite{Huang2006,HeLinMa2024}:

\[
|f_1(z)| = \frac{2}{\pi}\,\big|\sin\!\big(\pi D(z)\big)\big|, 
\qquad 
\arg f_1(z) = -\pi D(z).
\]
Through this slow modulation, the effective spatial dependence of the nonlinear
interaction is smoothly shaped along the propagation direction. Since the side
lobes of the sinc-type phasematching response arise from a uniform interaction
over a finite length, this shaping suppresses those side lobes and yields a
smooth, approximately Gaussian phasematching envelope.

\medskip
Apodization may thus be viewed as convolving the uniform-grating response with a slowly varying
envelope function. In the limit where the envelope is chosen to be Gaussian, the resulting
phasematching function is well-approximated by
\[
\phi_\text{apo}\!\left(\frac{\Delta k L}{2}\right)
\propto 
\exp\!\left[-\frac{(\Delta k L)^2\sigma^2}{2}\right],
\]
where \( \sigma \) characterizes the effective width of the nonlinear interaction region.
This Gaussian-like phasematching function suppresses side lobes and yields smoother JSAs at low
gain, which is the apodized configuration used in the main text.

\medskip
In summary, the unapodized case naturally yields the standard sinc-type phasematching function, while controlled apodization enables the nonlinear
profile to be engineered such that the resulting phasematching response becomes smooth and approximately Gaussian. These two limiting cases form the basis for
the comparison performed in Sec.~\ref{sec:simulation} of the main text.

\section{Other Group-velocity Configuration}
\label{app:other_gv_config}

\begin{figure*}[!htbp]
	\centering
	\includegraphics[width=\textwidth,keepaspectratio]{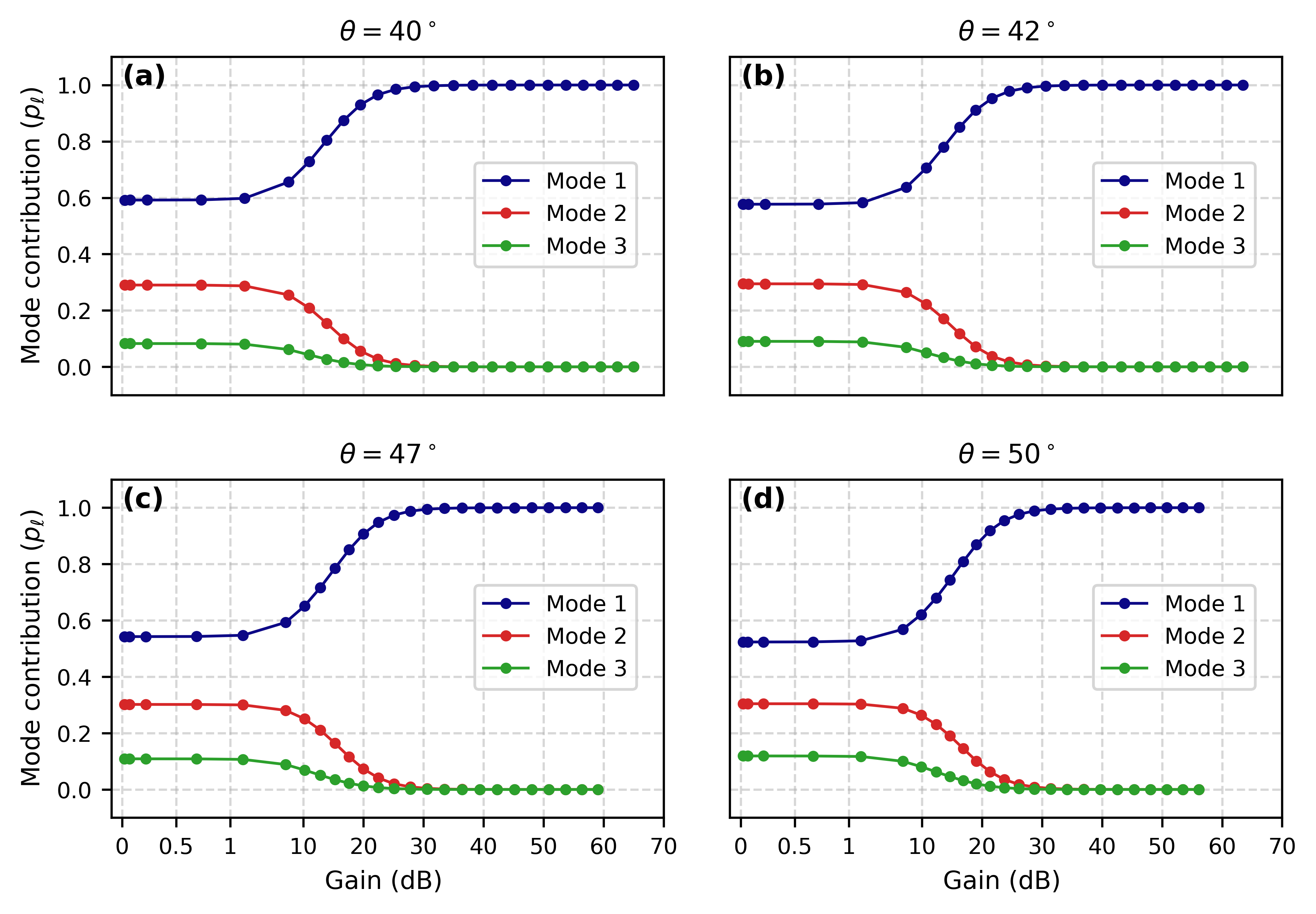} 
	\caption{In these figures, for a broadband pump in an unapodized WG (sinc-type phase matching), the mode contribution $p_{\ell}$ of the first three Schmidt modes is plotted versus gain for (a) $\theta=40^\circ$, (b) $\theta=42^\circ$, (c) $\theta=47^\circ$, and (d) $\theta=50^\circ$.}
	\label{fig:mode_contribution_sinc_40_42_47_50} 
\end{figure*}
As discussed in the main text, the dispersion configuration with $\theta = 45^\circ$ exhibits a characteristic high-gain suppression of higher-order Schmidt modes. In this Appendix, we show that the same behavior persists for nearby dispersion angles, specifically $\theta = 40^\circ$, $42^\circ$, $47^\circ$, and $50^\circ$, all of which satisfy the group-velocity ordering $v_g^{(s)} < v_g^{(p)} < v_g^{(i)}$.

Figure~\ref{fig:mode_contribution_sinc_40_42_47_50} shows the gain dependence of the mode contributions $p_\ell$ for the first three Schmidt modes under these four dispersion configurations. In all cases, the contribution of the first Schmidt mode increases rapidly with gain, while the second and third mode contributions decrease correspondingly.

This behavior mirrors the dynamics observed for the $\theta = 45^\circ$ case and confirms that the effect is not tied to a finely tuned symmetric point. Instead, it consistently appears whenever the pump group velocity lies between those of the signal and idler. As explained in Sec.~\ref{subsec:gain_dynamics}, this arises from the gain-dependent interference mechanism driven by the opposite-sign walk-off phases of the signal and idler fields. The results presented here therefore support the conclusion that the underlying mechanism is robust to small variations in the phasematching angle and is dictated primarily by the qualitative ordering of group velocities rather than the precise value of~$\theta$.

\nocite{*}

\bibliography{paper}

\end{document}